\documentclass[12pt,eqsecnum]{revtex4}
\begin{document}

\title{\Large BV formulation of higher form gauge theories in a superspace}

\author{ Sudhaker Upadhyay\footnote {e-mail address: sudhakerupadhyay@gmail.com}}
\author{ Bhabani Prasad Mandal\footnote{e-mail address:
bhabani.mandal@gmail.com}}

\affiliation { Department of Physics, \\
Banaras Hindu University, \\
Varanasi-221005, INDIA.  }
 
\begin{abstract}
We discuss the extended BRST and anti-BRST symmetry (including shift symmetry) in the  Batalin-Vilkovisky 
(BV) formulation for  two and three form gauge theories. Further we develop the superspace 
formulation for the BV actions for these theories. We show that the extended BRST invariant BV action for 
these theories can be written manifestly  covariant manner in a superspace with one Grassmann coordinate. On the
hand a superspace with two Grassmann coordinates are required for a manifestly covariant formulation 
of the extended BRST and extended anti-BRST invariant BV actions for higher form gauge theories.
\end{abstract}
\maketitle


\section{\large Introduction}	
Two and higher form gauge theories play an important role in the different branches of physics  \cite{green, pol}.
Low energy excitations in string theories contain states described by antisymmetric tensor fields \cite{h,i}. Various 
supergravity models are described in terms of antisymmetric tensor fields.  The  Abelian rank-2
tensor field plays crucial role in the study of classical string theories \cite{a}, in the theory of
vortex motion in an irrotational, incompressible fluid \cite{b,c}, in the 
dual formulation of the Abelian Higgs model \cite{d,e}, in   studying supergravity multiplets 
\cite{g} and in anomaly cancellation of certain superstring theories \cite{degu}.
The BRST symmetry is a fundamental tool for quantizing such theories
 and studying the renormalizability, unitarity and other aspects of different
gauge theories \cite{brst,tyu,ht,wei}. 
On the other hand BV formulation is known to be one of the most powerful method of quantizing different 
gauge field theories, supergravity theories and topological field theories in 
Lagrangian formulation \cite{ht,wei,bv,bv1,bv2,subm}. A superspace formalism for the
BV action of 1-form gauge theories has been studied \cite{ad, ba}.
It has been shown  how an extended BRST and extended
anti-BRST invariant formulation
(including some shift symmetry) of the BV action for these theories \cite{ad,ba,fk}, naturally leads to the 
proper identification
of the antifields through equations of motion of auxiliary field variables. Recently, this formulation 
has been extended for higher derivative theories \cite{fk}. We intend to extend such formulation 
beyond 1-form gauge theories. 

In this present article we discuss the BV formulation of extended BRST and extended anti-BRST invariant
higher form gauge theories. The extended  BRST and extended anti-BRST symmetry includes the shift 
symmetry of the fields. We further consider the superspace formulation of BV actions
 for  2-form and 3-form gauge theories. We show that the gauge-fixed Lagrangian density for such theories can be
described in the superspace formulation in the extended BRST invariant manner by considering
one Grassmann coordinate  $\theta$. On the other hand for manifestly extended BRST and extended anti-BRST
invariant formulation  of these theories, a superspace with two Grassmann coordinates is required. 

The paper is organized as the follows. We study the BV formalism in a superspace for 2-form gauge theory  in
section  II. A superspace formulations for 3-form gauge theories 
are discussed in Sec III. Conclusions are drawn in the last section.

\section{\large BV formulation of 2-form gauge theory in superspace}

We intend to discuss the BV formulation of 2-form gauge theories in a suitably constructed
superspace in this section.
In particular, we consider some shift symmetry and the usual BRST symmetry 
to construct an extended BRST invariant BV action.
Further, we develop an extended BRST invariant superspace formulation for such theory.
The extended anti-BRST symmetry for this BV  is also developed. Using all these formulation 
we finally construct the extended BRST and extended BRST and extended anti-BRST invariant BV action
in a superspace. 

\subsection{\large Shift symmetry and an extended BRST invariant BV action }
We start with the classical Lagrangian density for four dimensional Abelian   
rank-2 antisymmetric tensor field ($B_{\mu\nu}$) theory   
as
\begin{equation}
{\cal L}_0=\frac{1}{12}F_{\mu \nu \rho}F^{\mu \nu \rho},\label{kin}
\end{equation}
where the field-strength tensor ($F_{\mu \nu \rho}$) is defined as $F_{\mu \nu \rho}\equiv \partial_\mu 
B_{\nu\rho}+\partial_\nu B_{\rho\mu}+\partial_\rho 
B_{\mu\nu}.$  This  Lagrangian density is invariant under the following gauge transformation 
\begin{eqnarray}
\delta 
B_{\mu\nu}=\partial_{\mu}\zeta_{\nu} -\partial_{\nu}\zeta_{\mu},  
\end{eqnarray}
where $\zeta_{\mu}(x)$ an arbitrary vector field. 

To quantize this theory using BRST technique, it is necessary to introduce two
anticommuting vector fields $\rho_{\mu}$ and $\tilde\rho_{\mu}$, 
a commuting vector field $\beta_{\mu}$, two anticommuting scalar fields $\chi$ and $\tilde\chi$, 
and the commuting scalar fields $\sigma, \varphi$ and $ \tilde\sigma $ \cite{ht}.
The BRST invariant effective Lagrangian density for this theory in a covariant gauge is then given by
\begin{equation} 
{\cal L}_{eff} ={\cal L}_0 + {\cal L}_{gf}, \label{act}
\end{equation}
where the  gauge-fixing and ghost part of the Lagrangian density is given as \cite{sm}
\begin{eqnarray}
{\cal L}_{gf}&=& -i\partial_\mu\tilde\rho_\nu (\partial^\mu\rho^\nu -
\partial^\nu\rho^\mu )+\partial_\mu\tilde\sigma\partial^\mu\sigma +\beta_\nu(\partial_\mu B^{
\mu\nu} + k_1\beta^\nu -\partial^\nu\varphi)\nonumber\\ 
&-& i\tilde\chi\partial_\mu\rho^\mu -i\chi (\partial_\mu\tilde\rho^\mu -
k_2\tilde\chi), \label{gfix}
\end{eqnarray}
 $k_1$ and $k_2$ are arbitrary gauge parameters.
This effective theory is then invariant under following  BRST  
transformation:
\begin{eqnarray}
s_b B_{\mu\nu} &=& (\partial_\mu\rho_\nu -\partial_\nu\rho_\mu), \  
s_b\rho_\mu  =  -i\partial_\mu\sigma,     \ s_b\sigma 
= 0,\ s_b\tilde\rho_\mu  = i\beta_\mu, \nonumber\\
s_b\beta_\mu &=& 0, \ \ s_b\tilde\sigma  = -\tilde\chi,  \ \ \ 
s_b\tilde\chi =0, \ \
s_b\varphi  =  \chi,  \ \ s_b\chi =0.\label{sym}
\end{eqnarray}
Now, we extend this symmetry using shift symmetry in BV formulation. 

In the BV formalism, the gauge-fixing and ghost part of the Lagrangian density   
is generally expressed in terms of BRST variation of a gauge-fixed fermion.
It is straightforward to write the ${\cal L}_{gf}$ given in Eq. (\ref{gfix})
in terms of gauge-fixed fermion $\Psi$ 
as   
\begin{equation}
{\cal L}_{gf}=s_b \Psi,
\end{equation}
where the expression for $\Psi$ is 
\begin{equation}
\Psi =-i [\tilde\rho_\nu (\partial_\mu B^{\mu\nu} -k_1 \beta^\nu )
+\tilde\sigma\partial_\mu\rho^\mu +
\varphi (\partial_\mu\tilde\rho^\mu - k_2 \tilde\chi) ].\label{gff}
\end{equation}
To obtain the extended BRST invariant BV action for Abelian rank-2 tensor field theory we consider 
the following shifted  Lagrangian density \cite{al}
\begin{eqnarray}
\bar{\cal L}_{gf}&=& {\cal L}_{gf} (B_{\mu\nu}-\bar B_{\mu\nu},\rho_\mu 
-\bar\rho_\mu, \tilde\rho_{\mu}-\bar{
\tilde\rho}_{\mu}, \sigma_\mu - \bar{\sigma}_\mu, \tilde\sigma_{\mu}-\bar{\tilde\sigma}_{\mu
}, \beta_\mu - \bar\beta_\mu, \chi-\bar{\chi}, \tilde\chi - \bar{\tilde\chi},
\varphi -\bar\varphi )\nonumber\\
&=& -i(\partial_\mu\tilde\rho_\nu -\partial_\mu\bar{\tilde\rho}_\nu )(
\partial^\mu\rho^\nu -\partial^\mu\bar{\rho}^\nu -
\partial^\nu\rho^\mu +\partial^\nu\bar{\rho}^\mu )+(
\partial_\mu\tilde\sigma -\partial_\mu\bar{\tilde\sigma})(\partial^\mu\sigma -
\partial^\mu\bar\sigma )\nonumber\\
 &+&(\beta_\nu -\bar{\beta}_\nu ) (\partial_\mu B^{\mu\nu} -\partial_\mu \bar {B}^{\mu\nu}+ 
k_1\beta^\nu -k_1\bar{\beta}^\nu -\partial^\nu\varphi +\partial^\nu\bar{\varphi} )\nonumber\\
&-& i(\tilde\chi -\bar{\tilde\chi})(\partial_\mu\rho^\mu -\partial_\mu\bar{\rho}^\mu )
-i(\chi -\bar{\chi} ) (\partial_\mu\tilde\rho^\mu -\partial_\mu\bar{\tilde\rho}^\mu - k_2
\tilde\chi +k_2\bar{\tilde\chi}),
\end{eqnarray}
which coincides with  ${\cal L}_{gf}$ in Eq. (\ref{gfix}) when all the bar fields vanish. 
Here we notice that the  above Lagrangian density is invariant under the BRST transformation 
 (\ref{sym}) for the   fields $\Phi -\bar\Phi$, where $\Phi $ and $\bar\Phi  $ are generic notation for all
fields and shifted fields respectively. 
But in addition it is also invariant under the shift symmetry 
\begin{eqnarray}
s_b \Phi (x)= \alpha (x),\ \ s_b \bar\Phi (x)&=& \alpha (x).
\end{eqnarray}
  The BRST symmetry along with this shift symmetry form the extended BRST symmetry.
  The extended BRST transformation is then compactly written as
\begin{eqnarray}
s_b \Phi (x)= \alpha (x),\ \ s_b \bar\Phi (x)&=& \alpha (x)-\beta (x).
\end{eqnarray}
Here $\beta (x)$ represents the original BRST transformation of collective fields $\Phi$,
whereas $\alpha (x)$ corresponds to the shift transformation corresponding the collective fields $\Phi$.
This local shift symmetry needs to be gauge-fixed, which leads to an additional BRST symmetry \cite{ba}.      
The extended BRST symmetry transformation 
for the fields involved in the theory can be written explicitly as 
\begin{eqnarray}
s_b B_{\mu\nu} &=& \psi_{\mu\nu},\ \ s_b {\bar B}_{\mu\nu}= \psi_{\mu\nu}-
(\partial_\mu\rho_\nu -\partial_\mu{\bar \rho}_\nu-\partial_\nu\rho_\mu
+\partial_\nu{\bar\rho}_\mu),\ s_b\rho_\mu  =  \epsilon_\mu, \nonumber\\
 s_b\bar{\rho}_\mu &=& \epsilon_\mu +i\partial_\mu\sigma -i
\partial_\mu{\bar\sigma}, \ \
s_b\tilde\rho_\mu  = \xi_\mu,\  s_b\bar{\tilde\rho}_\mu =\xi_\mu -i\beta_\mu +i\bar{\beta}_\mu,\nonumber\\
 s_b\sigma  &=&  \varepsilon,\ \ s_b\bar\sigma = \varepsilon,\
s_b\beta_\mu  =  \eta_\mu,\ \ s_b\bar{\beta}_\mu = \eta_\mu, \
s_b\tilde\sigma  = \psi,\ \ s_b\tilde\chi =\eta,\nonumber\\ 
 s_b\bar{\tilde\sigma} &=& \psi +\tilde\chi -\bar{\tilde\chi},\ \ s_b\bar{\tilde\chi} = \eta,\ \ s_b\bar\varphi =
\phi- \chi +\bar\chi,\ 
 s_b\chi  = \Sigma,\nonumber\\
s_b\varphi &=& \phi,\ \ s_b\bar\chi =\Sigma,\
 s_b \xi_i=0,\ \xi_i\equiv [\psi_{\mu\nu}, \epsilon_\mu, \xi_\mu, \varepsilon, \eta_\mu, \psi, \eta, \phi, \Sigma ],
\label{brs1} 
\end{eqnarray}
where the fields $\psi_{\mu\nu},$  $\epsilon_\mu,$ $\xi_\mu,$ $\varepsilon,$ $\eta_\mu,$ $\psi,$ $\eta,$ 
$\phi$ and $ \Sigma$ are introduced as ghost fields associated with the shift symmetry for the fields 
$B_{\mu\nu},$ $\rho_\mu,$ $\tilde\rho_\mu,$ $\sigma,$ $\beta_\mu,$ $\tilde\sigma,$ $\tilde\chi,$ $\varphi$
and $\chi$ respectively.
Further, we  add antighosts fields $B_{\mu\nu}^\star,$ $\rho_{\mu}^\star,$ 
$\tilde{\rho}_{\mu}^\star,$ $\sigma^\star,$ $\tilde{\sigma}^\star,$ $\beta_\mu^\star,$ $\chi^\star,$
$\tilde{\chi}^\star$ and $ \varphi^\star $ with opposite parity, corresponding to
each of the fields with the following BRST transformations
\begin{eqnarray}
s_b B_{\mu\nu}^\star &=&L_{\mu\nu},\ \ 
s_b \rho_{\mu}^\star = M_\mu,\ \
s_b \tilde{\rho}_{\mu}^\star  =  \bar{M}_\mu,\ \
s_b \sigma^\star = N, \nonumber\\
s_b \tilde{\sigma}^\star &=& \bar N, \ \
s_b \beta_\mu^\star = S_\mu, \ \
s_b \chi^\star  =  O,\ \
s_b \tilde{\chi}^\star = \bar O, \nonumber\\
 s_b \varphi^\star  & =& T,\ \ s_b \chi_i =  0,
\chi_i \equiv [L_{\mu\nu},  M_\mu , \bar{M}_\mu, N, \bar N, S_\mu, O, \bar O, T], \label{brs2}  
\end{eqnarray}
where the fields $\chi_i$  are  the Nakanishi-Lautrup type auxiliary fields.

Now, if we gauge fix the shift symmetry by putting all the bar fields to zero we will be able to recover 
our  original 
theory.
This can be achieved by choosing the following gauge-fixed Lagrangian density
\begin{eqnarray}
\bar{\cal L}_{gf} &= & L_{\mu\nu}{\bar B}^{\mu\nu} -B_{\mu\nu}^\star (\psi^{\mu\nu}-
\partial^\mu\rho^\nu + 
\partial^\mu{\bar\rho}^\nu +\partial^\nu\rho^\mu
-\partial^\nu{\bar\rho}^\mu ) +{\bar M}_\mu{\bar\rho}^\mu \nonumber\\
&+&{\tilde \rho}_\mu^\star (\epsilon^\mu +i\partial^\mu\sigma -i\partial^\mu\bar\sigma ) +M_
\mu\bar{\tilde \rho}^\mu +\rho_\mu^\star (\xi^\mu-i\beta^\mu +i{\bar \beta}^\mu ) \nonumber\\
&+&N\bar\sigma -\sigma^\star \varepsilon  + \bar N\bar{\tilde\sigma} -{\tilde\sigma}^\star (
\psi-\tilde\chi +\bar{\tilde\chi}) +
\bar O\bar\chi +{\tilde\chi}^\star \Sigma +O\bar{\tilde\chi} \nonumber\\
&+&\chi^\star \eta+
T\bar\varphi -\varphi^\star (\phi -\chi+\bar\chi)
+S_\mu{\bar \beta}^\mu -\beta_\mu^\star \eta^\mu, \label{la}
\end{eqnarray}
which is invariant 
under the extended BRST symmetry transformations given in Eqs. (\ref{brs1}) and (\ref{brs2}).

Now, it is straightforward to check  that using equations of motion of 
auxiliary fields $\chi_i$   all the  bar fields  disappear from the above expression. The
 extended Lagrangian density $\bar{\cal L}_{gf}$ then can be cast in the following form:
\begin{eqnarray}
\bar{\cal L}_{gf} &= & -B_{\mu\nu}^\star (\psi^{\mu\nu}-
\partial^\mu\rho^\nu +\partial^\nu\rho^\mu )
 +{\tilde \rho}_\mu^\star (\epsilon^\mu +i\partial^\mu\sigma )
+ \rho_\mu^\star (\xi^\mu-i\beta^\mu )\nonumber\\
&-& \sigma^\star \epsilon   -{\tilde\sigma}^\star (\psi-\tilde\chi ) 
+{\tilde\chi}^\star \Sigma   
 + \chi^\star \eta 
-\varphi^\star (\phi -\chi )
-\beta_\mu^\star \eta^\mu. \label{per}
\end{eqnarray}
If the gauge-fixed  fermion $\Psi$ depends only on the original fields, then a
general gauge-fixing Lagrangian density 
for Abelian rank-2 antisymmetric tensor field with  original BRST symmetry 
will have the following form
\begin{eqnarray}
{\cal L}_{gf} &= &s_b \Psi =s_b B_{\mu\nu}\frac{\delta\Psi}{\delta B_{\mu\nu}}+
s_b\rho_\mu \frac{\delta\Psi}{\delta\rho_\mu}+ s_b\tilde{\rho}_\mu \frac{\delta\Psi}
{\delta\tilde{\rho}_\mu} +
s_b\sigma \frac{\delta\Psi}{\delta\sigma}\nonumber\\
&+&s_b\tilde\sigma \frac{\delta\Psi}{\delta
\tilde\sigma} +s_b\beta_\mu \frac{\delta\Psi}{\delta\beta_\mu}+
s_b\chi \frac{\delta\Psi}{\delta\chi}+s_b\tilde\chi \frac{\delta\Psi}{\delta\tilde\chi}
+s_b\varphi \frac{\delta\Psi}{\delta\varphi}, \nonumber\\
&= & \psi_{\mu\nu}\frac{\delta\Psi}{\delta B_{\mu\nu}}+
\epsilon _\mu \frac{\delta\Psi}{\delta\rho_\mu}+ \xi_\mu \frac{\delta\Psi}
{\delta\tilde{\rho}_\mu} +
\varepsilon \frac{\delta\Psi}{\delta\sigma}\nonumber\\
&+&\psi  \frac{\delta\Psi}{\delta
\tilde\sigma} +\eta_\mu \frac{\delta\Psi}{\delta\beta_\mu}+
\Sigma  \frac{\delta\Psi}{\delta\chi}+\eta \frac{\delta\Psi}{\delta\tilde\chi}
+\phi \frac{\delta\Psi}{\delta\varphi}.
\end{eqnarray}
Using the properties of the fields  the above gauge-fixed
Lagrangian density can  further be expressed as  
\begin{eqnarray}
{\cal L}_{gf} &= & -\frac{\delta\Psi}{\delta B_{\mu\nu}}\psi_{\mu\nu}+
 \frac{\delta\Psi}{\delta\rho_\mu}\epsilon _\mu + \frac{\delta\Psi}
{\delta\tilde{\rho}_\mu} \xi_\mu -
 \frac{\delta\Psi}{\delta\sigma}\varepsilon\nonumber\\
&-& \frac{\delta\Psi}{\delta
\tilde\sigma}\psi -\frac{\delta\Psi}{\delta\beta_\mu}\eta_\mu +
\frac{\delta\Psi}{\delta\chi}\Sigma + \frac{\delta\Psi}{\delta\tilde\chi}\eta
-\frac{\delta\Psi}{\delta\varphi}\phi. \label{psi} 
\end{eqnarray}
Now, the total Lagrangian density ${\cal L}_{T} =  {\cal L}_0+
{\cal L}_{gf}+\bar{\cal L}_{gf}$  is then given as 
\begin{eqnarray}
{\cal L}_{T}&=&
  \frac{1}{12}F_{\mu \nu \rho}F^{\mu \nu \rho} + B_{\mu\nu}^\star (
\partial^\mu\rho^\nu -\partial^\nu\rho^\mu )
 +i{\tilde \rho}_\mu^\star \partial^\mu\sigma 
 -i\rho_\mu^\star \beta^\mu 
 + {\tilde\sigma}^\star \tilde\chi
+\varphi^\star \chi \nonumber\\
&-&\left(B_{\mu\nu}^\star +\frac{\delta\Psi}{\delta B^{\mu\nu}}\right)\psi^{\mu\nu}
+\left({ \rho}_\mu^\star +\frac{\delta\Psi}{\delta\tilde{\rho}^\mu}\right)\xi^\mu  
 +\left({\tilde \rho}_\mu^\star +\frac{\delta\Psi}{\delta\rho^\mu}\right)\epsilon^\mu 
-\left(\sigma^\star +\frac{\delta\Psi}{\delta\sigma}\right)\varepsilon \nonumber\\
&-& \left (\tilde{\sigma}^\star +\frac{\delta\Psi}{\delta\tilde\sigma} \right )\psi
+ \left (\tilde{\chi}^\star +\frac{\delta\Psi}{\delta\chi} \right )\Sigma 
+ \left ({\chi}^\star +\frac{\delta\Psi}{\delta\tilde\chi} \right )\eta 
-\left (\varphi^\star +\frac{\delta\Psi}{\delta\varphi} \right )\phi\nonumber\\
&-& \left (\beta_\mu^\star +\frac{\delta\Psi}{\delta\beta^\mu} \right )\eta^\mu,
\end{eqnarray}
where we have used the Eqs. (\ref{kin}), (\ref{per}) and (\ref{psi}).
Integration over
ghost fields  associated with the shift symmetry leads to the following identification  
\begin{eqnarray}
B_{\mu\nu}^\star &=&-\frac{\delta\Psi}{\delta B^{\mu\nu}},\ \
{\tilde \rho}_\mu^\star =-\frac{\delta\Psi}{\delta\rho^\mu},\ \
{ \rho}_\mu^\star  = -\frac{\delta\Psi}{\delta\tilde{\rho}^\mu},\nonumber\\
\sigma^\star &=&-\frac{\delta\Psi}{\delta\sigma},\ \ 
\tilde{\sigma}^\star  = -\frac{\delta\Psi}{\delta\tilde\sigma},\ \ 
\tilde{\chi}^\star =-\frac{\delta\Psi}{\delta\chi},\nonumber\\
{\chi}^\star &=&-\frac{\delta\Psi}{\delta\tilde\chi},\ \
\beta_\mu^\star =-\frac{\delta\Psi}{\delta\beta^\mu}, \ \ 
\varphi^\star  = -\frac{\delta\Psi}{\delta\varphi}.
\end{eqnarray}
Using above equation and the gauge-fixed fermion given in Eq. (\ref{gff}),
 we obtain the antifields associated with this
theory as 
\begin{eqnarray}
B_{\mu\nu}^\star &=&-i\partial_\mu\tilde{\rho}_\nu,\ \
{\tilde \rho}_\mu^\star =-i\partial_\mu\tilde{\sigma},\ \ 
{ \rho}_\mu^\star  = i(\partial^\nu B_{\nu\mu} -k_1\beta_\mu
),\ \
\sigma^\star =0\nonumber\\
\tilde{\sigma}^\star &=&i\partial_\mu\rho^\mu,\  
\tilde{\chi}^\star =0,\  
{\chi}^\star  = -i k_2\varphi,\ 
\beta_\mu^\star =i k_1\tilde{\rho}_\mu,\  
\varphi^\star  = i(\partial_\mu\tilde{\rho}^\mu -k_2\tilde\chi). \label{antifield}
\end{eqnarray}
With these identification the total Lagrangian density reduces to the original 
theory for Abelian rank-2 tensor field given in Eq. (\ref{act}). 
Now, we are  able to write the  gauge-fixing part of the total Lagrangian density in terms of
the BRST variation of a generalized gauge-fixed fermion as follows 
\begin{eqnarray}
 {\cal L}_{gf} + \bar{\cal L}_{gf}&=&   s_b \left(B_{\mu\nu}^\star \bar B^{\mu\nu} +
\rho_{\mu}^\star \bar {\tilde\rho}^{\mu} +
\tilde\rho_{\mu}^\star \bar{ \rho}^{\mu} +\sigma^\star\bar\sigma +
\tilde\sigma^\star\bar{\tilde\sigma}+\beta_\mu^\star\bar\beta^\mu  \right.\nonumber\\
&+&\left. 
\chi^\star\bar{\tilde\chi} +\tilde\chi^\star\bar{\chi}
+\varphi^\star\bar\varphi
\right) 
 \equiv  {\cal L}_0+s_b \Phi^\star\bar\Phi,
\end{eqnarray}
where the fields $\Phi^\star$ and $\bar\Phi$ are the generic notation
for antifields and corresponding shifted fields respectively. As 
expected the ghost number of   $ \Phi^\star\bar\Phi$   is $-1$. We recover the BV action for
Abelian 2-form gauge theory with the identification of antifields given in Eq. (\ref{antifield}). 

Next, we construct the superspace formulation of such BRST invariant extended theory. 
\subsection{\large  Extended BRST invariant superspace formulation}   
In this section we develop a superspace
formalism of the extended BRST invariant theory discussed in
 the previous section.  For this purpose we consider a superspace with 
coordinates $(x^\mu, \theta)$ where  $\theta$ is fermionic coordinate. 
In such a superspace the ``superconnection" $2$-form can be written as \cite{bt}
\begin{equation}
\omega^{(2)} =\frac{1}{2 !}{\cal B}_{\mu\nu}(x, \theta ) (dx^\mu\wedge dx^\nu ) +
 {\cal M}_{\mu}(x, \theta ) (dx^\mu\wedge d\theta ) + {\cal N} (x, \theta ) (d\theta\wedge d\theta ), 
\end{equation}
where $d $ is an exterior derivative and is defined as $d=dx^\mu\partial_\mu +d\theta \partial_\theta$.
The requirement for super curvature (field
strength   $F^{(3)}= d\omega^{(2)}$) to vanish
along the $\theta$ direction restricts 
the component of the superfields  to have   following form
\begin{eqnarray}
{\cal B}_{\mu\nu}(x, \theta ) &=& B_{\mu\nu} (x) +\theta  (s_bB_{\mu\nu}),\nonumber\\
{\cal M}_{\mu}(x, \theta ) &=& \rho_\mu (x) +\theta (s_b\rho_\mu),\nonumber\\
 {\cal N} (x, \theta ) &=& \sigma (x) +\theta (s_b \sigma). 
\end{eqnarray}
Similarly, we  define all the superfields corresponding to each fields  involved in extended BV action  as 
\begin{eqnarray} 
{\cal B}_{\mu\nu}(x, \theta ) &=& B_{\mu\nu} (x) +\theta \psi_{\mu\nu},\ \ {\cal M}_{\mu}(x, \theta )  =  \rho_{\mu} 
(x) +\theta \epsilon_\mu,\nonumber\\
\bar{{\cal B}}_{\mu\nu}(x, \theta ) &=& \bar{B}_{\mu\nu} (x) +\theta (
\psi_{\mu\nu}-\partial_\mu\rho_\nu + \partial_\mu\bar{\rho}_\nu +
\partial_\nu\rho_\mu -\partial_\nu\bar{\rho}_\mu ),\nonumber\\
\bar{{\cal{M}}}_{\mu}(x, \theta ) &=& \bar{\rho}_{\mu} (x) +\theta (
\epsilon_\mu -i\partial_\mu\sigma +i\partial_\mu\bar\sigma ),\nonumber\\
{\cal{N}}(x, \theta ) &=& \sigma (x) +\theta 
\varepsilon,\ \
\bar{{\cal{N}}}(x, \theta )  =  \bar{\sigma}(x) +\theta \varepsilon,  \nonumber\\
\tilde{\cal M}_{\mu}(x, \theta ) &=& \tilde\rho_{\mu} (x) +\theta \xi_\mu,\ \
\bar{\tilde{\cal M}}_{\mu}(x, \theta )  =  \bar{\tilde\rho}_{\mu} (x) +\theta (
\xi_\mu -i\beta_\mu +i\bar\beta_\mu ), \nonumber\\
{\cal S}_{\mu}(x, \theta ) &=& \beta_{\mu} (x) +\theta \eta_\mu,\ \ 
\bar{\cal S}_{\mu}(x, \theta )  =  \bar\beta_{\mu} (x) +\theta \eta_\mu, \nonumber\\
\tilde{\cal N}(x, \theta ) &=& \tilde\sigma (x) +\theta \psi,\ \
\bar{\tilde{\cal N}}(x, \theta )  = \bar{\tilde\sigma} (x) +\theta (\psi -\tilde
\chi +\bar{\tilde\psi}), \nonumber\\
{\cal{O}}(x, \theta ) &=& \chi (x) +\theta 
\Sigma,\ \
\bar{\cal{O}}(x, \theta )  =  \bar\chi (x) +\theta 
\Sigma, \nonumber\\
\tilde{\cal{O}}(x, \theta ) &=& \tilde\chi (x) +\theta 
\eta, \ \
\bar{\tilde{\cal{O}}}(x, \theta )  =  \bar{\tilde\chi} (x) +\theta 
\eta, \nonumber\\
{\cal{T}}(x, \theta ) &=& \varphi (x) +\theta 
\phi, \ \
\bar{\cal{T}}(x, \theta ) =  \bar\varphi (x) +\theta 
(\phi -\chi +\bar\chi ).\label{sufi}
\end{eqnarray} 
The super antifields also have the following form
\begin{eqnarray}
\bar{\cal B}_{\mu\nu}^\star &=& B_{\mu\nu}^\star  +\theta L_{\mu\nu}, \  
\bar{{\cal M}_{\mu}}^\star = \rho_{\mu}^\star  +\theta M_{\mu},\
\bar{\tilde{\cal M}}_{\mu}^\star  =  \tilde\rho_{\mu}^\star  +\theta \bar M_{\mu},\nonumber\\
\bar{{\cal S}_{\mu}}^\star &=& \beta_{\mu}^\star  +\theta S_{\mu}, \ \
\bar{{\cal N}}^\star  =  \sigma^\star  +\theta N,\ \
\bar{\tilde{\cal N}}^\star = \tilde\sigma^\star  +\theta \bar N,\nonumber\\
\bar{{\cal O}}^\star &=& \chi^\star  +\theta O,\ \
\bar{{\cal T}}^\star = \varphi^\star  +\theta T, \ \
\bar{\tilde{\cal O}}^\star  = \tilde\chi^\star  +\theta \bar O.\label{antisufi}
\end{eqnarray}
The Eqs. (\ref{sufi}) and (\ref{antisufi}) enable us to write   
\begin{eqnarray}
&&\frac{\delta}{\delta\theta}\bar{\cal B}_{\mu\nu}^\star \bar{\cal B}^{\mu\nu}
 =  L_{\mu\nu}\bar {B}^{\mu\nu} -B_{\mu\nu}^\star  (\psi^{\mu\nu} -\partial^\mu\rho^\nu +
\partial^\mu\rho^\nu
+\partial^\nu\rho^\mu -\partial^\nu\bar{\rho}^\mu ),
\nonumber\\
&&\frac{\delta}{\delta\theta}\bar{\tilde{\cal M}}_{\mu}^\star  \bar{\cal M}^{\mu}
 =  \bar M_{\mu}\bar{\rho}^\mu +\tilde\rho_{\mu}^\star  (\epsilon ^{\mu}+
i\partial^\mu\sigma -i\partial^\mu\bar\sigma ),
\nonumber\\
&&\frac{\delta}{\delta\theta}\bar{\tilde{\cal M}}_{\mu} \bar{\cal M}^{\mu\star}
=  M_{\mu}\bar{\tilde\rho}^\mu +\rho_{\mu}^\star  (\xi^{\mu}-
i\beta^\mu -i\bar\beta^\mu ),
\nonumber\\
&&\frac{\delta}{\delta\theta}\bar{\cal N}^\star  \bar{\cal N}
= N\bar{\sigma} -\sigma^\star  \varepsilon, 
\ \
\frac{\delta}{\delta\theta}\bar{\tilde{\cal N}}^\star  \bar{\tilde{\cal N}}
 =  \bar N\bar{\tilde\sigma} -\tilde\sigma^\star  (\psi-\tilde\chi +\bar{\tilde\chi}),
\nonumber\\
&&\frac{\delta}{\delta\theta}\bar{\tilde{\cal O}}^\star  \bar{{\cal O}}
= \bar O\bar{\chi} +\tilde\chi^\star  \Sigma,
 \ \
\frac{\delta}{\delta\theta}\bar{\tilde{\cal O}}  \bar{{\cal O}}^\star
 =  O\bar{\tilde\chi} +\chi^\star  \eta,
\nonumber\\
&&\frac{\delta}{\delta\theta}\bar{\cal T}^\star  \bar{\cal T}
=  T\bar{\varphi} -\varphi^\star ( \phi -\chi +\bar\chi ),
\ \
\frac{\delta}{\delta\theta}\bar{\cal S}_\mu^\star  \bar{\cal S}^\mu
 =   S_\mu\bar{\beta}^\mu -\beta_\mu^\star  \eta^\mu.
\end{eqnarray} 
Then the gauge-fixed   
Lagrangian density for shift symmetry  given in Eq. (\ref{la}) can be written  in the 
superspace formulation as
\begin{eqnarray}
\bar{\cal L}_{gf} &=& \frac{\delta}{\delta\theta}\left[\bar{\cal B}_{\mu\nu}^\star \bar{\cal 
B}^{\mu\nu} +\bar{\tilde{\cal M}}_{\mu}^\star  \bar{\cal M}^{\mu}+\bar{\tilde{\cal 
M}}_{\mu} \bar{\cal M}^{\mu\star}+\bar{\cal N}^\star  \bar{\cal N}+
\bar{\tilde{\cal N}}^\star  \bar{\tilde{\cal N}}+
\bar{\tilde{\cal O}}^\star  \bar{{\cal O}}+\bar{\tilde{\cal O}}  \bar{{\cal O}}^
\star \right.\nonumber\\
&+&\left.\bar{\cal T}^\star  \bar{\cal T}+\bar{\cal S}_\mu^\star  \bar{\cal 
S}^\mu\right].\label{super}
\end{eqnarray}
The $\bar{\cal L}_{gf} $ remains invariant under the
extended BRST transformation as it belongs to the $\theta$ component of superfields. If the 
gauge-fixing fermion depends only on the original fields, then one can define the fermionic superfield  $\Gamma $ as
\begin{eqnarray}
{\Gamma  } &=&\Psi +\theta s_b \Psi,\nonumber\\ 
  &=&\Psi  + \theta \left[-\frac{\delta\Psi}{\delta B_{\mu\nu}}\psi_{\mu\nu}+
 \frac{\delta\Psi}{\delta\rho_\mu}\epsilon _\mu + \frac{\delta\Psi}
{\delta\tilde{\rho}_\mu} \xi_\mu -
 \frac{\delta\Psi}{\delta\sigma}\varepsilon\right.\nonumber\\
&-&\left. \frac{\delta\Psi}{\delta
\tilde\sigma}\psi -\frac{\delta\Psi}{\delta\beta_\mu}\eta_\mu +
\frac{\delta\Psi}{\delta\chi}\Sigma + \frac{\delta\Psi}{\delta\tilde\chi}\eta
-\frac{\delta\Psi}{\delta\varphi}\phi\right].
\end{eqnarray}
With these realization the original gauge-fixing Lagrangian density ${\cal L}_{gf} $ in the superspace
formalism can be expressed as 
\begin{equation}
{\cal L}_{gf} =\frac{\delta\Gamma }{\delta\theta}.\label{lagr}
\end{equation}
Further we notice that invariance  of ${\cal L}_{gf}$ under the extended BRST transformation is assured as it is
 $\theta$ component of fermionic superfield. Combining Eqs. (\ref{super}) and (\ref{lagr}) we can write the 
total Lagrangian density
 in this formalism as 
\begin{eqnarray}
 {\cal L}_{T} &=&{\cal L}_0 +{\cal L}_{gf} +\bar {\cal L}_{gf},\nonumber\\
& =& {\cal L}_0 + \frac{\delta}{\delta\theta}\left[\bar{\cal B}_{\mu\nu}^\star \bar{\cal 
B}^{\mu\nu} +\bar{\tilde{\cal M}}_{\mu}^\star  \bar{\cal M}^{\mu}+\bar{\tilde{\cal 
M}}_{\mu} \bar{\cal M}^{\mu\star}+\bar{\cal N}^\star  \bar{\cal N}+
\bar{\tilde{\cal N}}^\star  \bar{\tilde{\cal N}}+
\bar{\tilde{\cal O}}^\star  \bar{{\cal O}}+\bar{\tilde{\cal O}}  \bar{{\cal O}}^
\star \right.\nonumber\\
&+&\left. \bar{\cal T}^\star  \bar{\cal T}+\bar{\cal S}_\mu^\star  \bar{\cal 
S}^\mu +\Gamma\right].                                        
\end{eqnarray}
If one performs equations of motion of auxiliary fields and ghost fields associated with the shift symmetry, 
this  Lagrangian density reduces to the original BRST invariant Lagrangian density.
\subsection{\large Extended anti-BRST invariant BV action} 
In the previous subsections we have analyzed the extended BRST symmetry of the Abelian rank-2 antisymmetric tensor field
and have developed the corresponding superspace formulation. In this subsection we construct the extended 
anti-BRST invariant Lagrangian density for the same
theory. For that we start with the anti-BRST symmetry transformation ($s_{ab}$), under  which the
Lagrangian density for the 2-form gauge theory given in Eq. (\ref{act}), remains invariant, as 
\begin{eqnarray}
s_{ab} B_{\mu\nu} &=& (\partial_\mu\tilde\rho_\nu -\partial_\nu\tilde\rho_\mu),\  
s_{ab}\tilde\rho_\mu  =  -i\partial_\mu\tilde\sigma , \ s_{ab}
\tilde\sigma 
= 0,  \
s_{ab}\rho_\mu  = -i\beta_\mu,\nonumber\\
s_{ab}\beta_\mu &= &0,\ 
s_{ab}\sigma  = \chi,   \
s_{ab}\chi =0, \
s_{ab}\varphi  =  -\tilde\chi,  \ \ s_{ab}\tilde\chi =0.
\label{anbr}
\end{eqnarray}
We note that the above anti-BRST transformation does not  absolutely anticommute   with the 
BRST transformation given in Eq. (\ref{sym}) i.e. $\{s_b, s_{ab}\}\not =0$.
But one can achieve the absolutely anticommutativity of these by considering a Curci-Ferrari (CF) type
restriction \cite{cf} in this theory. We will emphasize these with more details in the case 
of Abelian 3-form gauge theory in the next subsection. 
 
The gauge-fixed {\it anti-fermion}  $\bar\Psi$ ( gauge-fixing fermion in case of anti-BRST transformation)
 for this theory is defined as
\begin{equation}
\bar\Psi =i\left[\rho_\nu (\partial_\mu B^{\mu\nu} +k_1 \beta^\nu )
-\sigma\partial_\mu\tilde\rho^\mu +
\varphi (\partial_\mu\rho^\mu + k_2 \chi)\right],
\end{equation}
to write  the gauge-fixing part of the Lagrangian density  in terms of anti-BRST variation of $\bar\Psi$ as
\begin{equation}
{\cal L}_{gf}=s_{ab} \bar\Psi= i s_{ab} [\rho_\nu (\partial_\mu B^{\mu\nu} +k_1 \beta^\nu )
-\sigma\partial_\mu\tilde\rho^\mu +
\varphi (\partial_\mu\rho^\mu + k_2 \chi)].
\end{equation}
Following the same procedure as in the case of BRST transformation, we demand that $s_{ab}(\Phi-\bar\Phi)$ 
reproduces the anti-BRST transformations of ordinary fields $(\Phi)$ for
the Abelian rank-2 antisymmetric tensor field theory mentioned in Eq. (\ref{anbr}), 
 which leads to the following transformations
\begin{eqnarray}
s_{ab} \bar {B}_{\mu\nu} &=& B_{\mu\nu}^\star,\ \ s_{ab} {B}_{\mu\nu}= B_{\mu\nu}^\star +
(\partial_\mu\tilde\rho_\nu -\partial_\mu\bar{ \tilde\rho}_\nu-\partial_\nu\tilde\rho_\mu
+\partial_\nu\bar{\tilde\rho}_\mu),\nonumber\\
s_{ab}\bar{\tilde{\rho}}_\mu &=& \tilde\rho_\mu^\star ,\ \ s_{ab}\tilde{\rho}_\mu = \tilde\rho_\mu^\star 
-i\partial_\mu\tilde\sigma +i
\partial_\mu\bar{\tilde\sigma}, \ s_{ab}\bar\rho_\mu =\rho_\mu^\star,\nonumber\\
 s_{ab}\rho_\mu &=&\rho_\mu^\star  -i\beta_\mu +i\bar{\beta}_\mu,
\ \
s_{ab}\bar{\tilde\sigma}  =  \tilde\sigma^\star,\   s_{ab}\tilde\sigma =\tilde\sigma^\star,
\ s_{ab}\bar\beta_\mu =  \beta_\mu^\star,\nonumber\\
 s_{ab}{\beta}_\mu &=& \beta_\mu^\star,\
s_{ab}\bar\sigma = \sigma^\star,\ \ s_{ab}{\sigma} =\sigma^\star -\chi +\bar{\chi}, \ s_{ab}\bar\chi =\chi^\star,
\nonumber\\
 s_{ab}{\chi}&=&\chi^\star, \
s_{ab}\bar\varphi  =  \varphi^\star,\ \ s_{ab}\varphi =\varphi^\star- \tilde\chi +\bar{
\tilde\chi},\  s_{ab}\bar{\tilde\chi} =\tilde\chi^\star, \nonumber\\
s_{ab}\tilde\chi &=&\tilde\chi^\star, \
 s_{ab}\xi^\star =0,\ \xi^\star \equiv [B_{\mu\nu}^\star, \tilde\rho_\mu^\star, \rho_\mu^\star, \tilde\sigma^\star, 
\beta_\mu^\star, \psi,
\sigma^\star, \chi^\star, \varphi^\star, \tilde\chi^\star].\label{ab}
\end{eqnarray}
The ghost fields associated with the shift symmetry have the following extended
anti-BRST transformations,
\begin{eqnarray}
s_{ab}\psi_{\mu\nu}&=& L_{\mu\nu},\ \ s_{ab}\epsilon_{\mu} = M_{\mu},\ \
s_{ab}\xi_{\mu} =\bar M_{\mu},\ \
s_{ab}\varepsilon = N,\nonumber\\
 s_{ab}\psi  &=& \bar N,\ 
s_{ab}\eta_{\mu} = S_{\mu},\ 
s_{ab}\Sigma  =  O,\   s_{ab}\eta = \bar O,\  
s_{ab}\phi =T,\nonumber\\
s_{ab}\bar M_{\mu} &=&O,\ \ s_{ab} \omega  =0,\
\omega \equiv [L_{\mu\nu}, M_{\mu}, N, \bar N, S_\mu, O, \bar O, T].
\end{eqnarray}
These transformations along with the transformations in Eq. (\ref{ab}) consist the
extended anti-BRST transformations under which the total Lagrangian density including 
the shift fields remains invariant.
These transformations will be helpful  to establish the results in superspace formulation of BV action
in the next section.
\subsection{\large  Extended BRST and anti-BRST invariant superspace formulation}
To write a Lagrangian density that is manifestly invariant  under the both extended
BRST  and   extended anti-BRST transformations we need to define a 
 superspace  with  two Grassmannian 
coordinates $\theta$ and $\bar\theta$.
All the superfields in this superspace are the function of $(x_\mu, \theta, \bar\theta)$. In this situation the 
``super connection" $2$-form ($\omega^{(2)}$) 
and field strength ($ F^{(3)}$) are defined as
\begin{eqnarray}
\omega^{(2)} &=&\frac{1}{2 !}{\cal B}_{\mu\nu}(x, \theta, \bar\theta ) (dx^\mu\wedge dx^\nu ) +
 {\cal M}_{\mu}(x, \theta, \bar\theta ) (dx^\mu\wedge d\theta ) + {\cal N} (x, \theta, \bar\theta ) 
(d\theta\wedge d\theta )\nonumber\\
&+& \tilde{\cal M}_{\mu}(x, \bar\theta, \bar\theta ) (dx^\mu\wedge d\bar\theta ) + \tilde{\cal N}
 (x, \bar\theta, \bar\theta ) (d 
\bar\theta\wedge d\bar\theta ) +{\cal T} (x, \bar\theta, \bar\theta )(d \theta\wedge d\bar\theta ),\\
 F^{(3)} &=&d \omega^{(2)},
\end{eqnarray} 
where the exterior derivative has the following structure $d=dx^\mu\partial_\mu +d\theta\partial_\theta +
d\bar \theta\partial_{\bar\theta}$.

The requirement of vanishing the field strength corresponding to the extended theory along the directions
$\theta$ and $\bar\theta$ produces the superfields to have the following forms 
\begin{eqnarray}
 {\cal B}_{\mu\nu}(x, \theta, \bar\theta ) &=& B_{\mu\nu} (x) +\theta \psi_{\mu\nu}
+\bar\theta ( B_{\mu\nu}^\star + 
\partial_\mu\tilde\rho_\nu -\partial_\mu\bar{ \tilde\rho}_\nu-\partial_\nu\tilde\rho_\mu
+\partial_\nu\bar{\tilde\rho}_\mu) \nonumber\\
 &+ &\theta\bar\theta [L_{\mu\nu} +i(\partial_\mu\beta_\nu -\partial_\nu\beta_\mu
-\partial_\mu\bar\beta_\nu +\partial_\nu\bar\beta_\mu )],\nonumber\\
 \bar{{\cal B}}_{\mu\nu}(x, \theta, \bar\theta ) &=& \bar{B}_{\mu\nu} (x) 
 +\theta (
\psi_{\mu\nu}-\partial_\mu\rho_\nu + \partial_\mu\bar{\rho}_\nu +
\partial_\nu\rho_\mu -\partial_\nu\bar{\rho}_\mu )+\bar\theta B_{\mu\nu}^\star +\theta\bar\theta L_{\mu\nu},
\nonumber\\
{\cal M}_{\mu}(x, \theta, \bar\theta ) 
&=&\rho_{\mu} (x) +\theta \epsilon_\mu +\bar\theta 
( \rho_\mu^\star  -i\beta_\mu +i\bar{\beta}_\mu ) +\theta\bar\theta M_{\mu},\nonumber\\
\bar{{\cal{M}}}_{\mu}(x, \theta, \bar\theta )&=& \bar{\rho}_{\mu} (x) 
 +\theta (\epsilon_\mu -i\partial_\mu\sigma +i\partial_\mu\bar\sigma ) +\bar\theta 
\rho_\mu^\star +\theta\bar\theta M_{\mu},\nonumber\\
{\cal{N}}(x, \theta, \bar\theta )&=&\sigma (x) +\theta 
\varepsilon  
  +\bar\theta(\sigma^\star +\chi -\bar\chi) +\theta\bar\theta N,\ 
 \bar{{\cal{N}}}(x, \theta, \bar\theta ) = \bar{\sigma}(x) +\theta \varepsilon  
+\bar\theta \sigma^\star +\theta\bar\theta N,\nonumber\\
 \tilde{\cal M}_{\mu}(x, \theta, \bar\theta ) &=& \tilde\rho_{\mu} (x) +\theta \xi_\mu +
\bar\theta (\tilde\rho_\mu^\star -i\partial_\mu\tilde\sigma +i
\partial_\mu\bar{\tilde\sigma}) +\theta\bar\theta (\bar M_\mu -i\partial_\mu\tilde\chi +
i\partial_\mu\bar{\tilde\chi} ),\nonumber\\
 \bar{\tilde{\cal M}}_{\mu}(x, \theta, \bar\theta )&=& \bar{\tilde\rho}_{\mu} (x) +\theta (
\xi_\mu -i\beta_\mu +i\bar\beta_\mu ) +\bar\theta\tilde\rho_\mu^\star +\theta\bar\theta \bar 
M_\mu,\nonumber\\
{\cal S}_{\mu}(x, \theta, \bar\theta )  
&=&\beta_{\mu} (x) +\theta \eta_\mu +
\bar\theta\beta_\mu^\star +\theta\bar\theta S_\mu, \
 \bar{\cal S}_{\mu}(x, \theta, \bar\theta )= \bar\beta_{\mu} (x) +\theta \eta_\mu 
+\bar\theta\beta_\mu^\star +\theta\bar\theta S_\mu,\nonumber\\
 \tilde{\cal N}(x, \theta, \bar\theta )&=& \tilde\sigma (x) +\theta \psi
 +\bar\theta\tilde\sigma^\star +\theta\bar\theta \bar N,\ 
\bar{\tilde{\cal N}}(x, \theta, \bar\theta )=\bar{\tilde\sigma} (x) 
+\theta (\psi -\tilde
\chi +\bar{\tilde\psi}) 
 +\bar\theta\tilde\sigma^\star +\theta\bar\theta \bar N,\nonumber\\ 
 {\cal{O}}(x, \theta, \bar\theta )&=& \chi (x) +\theta 
\Sigma   +\bar\theta\chi^\star +\theta\bar\theta O,\
\bar{\cal{O}}(x, \theta, \bar\theta ) = \bar\chi (x) 
  +\theta 
\Sigma   +\bar\theta\chi^\star +\theta\bar\theta O,\nonumber\\ 
 \tilde{\cal{O}}(x, \theta, \bar\theta )&=& \tilde\chi (x) +\theta 
\eta +\bar\theta\tilde\chi^\star +\theta\bar\theta \bar O,\ 
\bar{\tilde{\cal{O}}}(x, \theta, \bar\theta ) 
 = \bar{\tilde\chi} (x) +\theta 
\eta +\bar\theta\tilde\chi^\star +\theta\bar\theta \bar O,\nonumber\\
 {\cal{T}}(x, \theta, \bar\theta ) &=& \varphi (x) +\theta 
\phi +\bar\theta (\varphi^\star -\tilde\chi +\bar{\tilde\chi}) +\theta\bar\theta T,
\nonumber\\
 \bar{\cal{T}}(x, \theta, \bar\theta ) &=&\bar\varphi (x) +\theta 
(\phi -\chi +\bar\chi ) +\bar\theta\varphi^\star +\theta\bar\theta T.
\end{eqnarray}
From the structure of superfields, we calculate the following relations   
\begin{eqnarray}
&&\frac{1}{2}\frac{\delta}{\delta\bar\theta}\frac{\delta}{\delta\theta}\bar{\cal B}_{\mu\nu}
\bar{\cal B}^{\mu\nu}
 =  L_{\mu\nu}\bar {B}^{\mu\nu} -B_{\mu\nu}^\star  (\psi^{\mu\nu} -\partial^\mu\rho^\nu +
\partial^\mu\rho^\nu
+\partial^\nu\rho^\mu -\partial^\nu\bar{\rho}^\mu ),
\nonumber\\
 &&\frac{\delta}{\delta\bar\theta}\frac{\delta}{\delta\theta}\bar{\tilde{\cal M }}_{\mu} 
\bar{\cal M}^{\mu} = \bar M_{\mu}\bar{\rho}^\mu +\tilde\rho_{\mu}^\star  (\epsilon ^{\mu}+
i\partial^\mu\sigma -i\partial^\mu\bar\sigma )
+ M_{\mu}\bar{\tilde\rho}^\mu +\rho_{\mu}^\star  (\xi^{\mu}-
i\beta^\mu +i\bar\beta^\mu ),
\nonumber\\
&&\frac{1}{2}\frac{\delta}{\delta\bar\theta}\frac{\delta}{\delta\theta}\bar{\cal N}
 \bar{\cal N}
=  N\bar{\sigma} -\sigma^\star  \varepsilon, 
\ \ \ \frac{1}{2}\frac{\delta}{\delta\bar\theta}\frac{\delta}{\delta\theta}\bar{\tilde{\cal N
}} \bar{\tilde{\cal N}}
= \bar N\bar{\tilde\sigma} -\tilde\sigma^\star  (\psi-\tilde\chi +\bar{\tilde\chi}),
\nonumber\\
&& \frac{\delta}{\delta\bar\theta}\frac{\delta}{\delta\theta}\bar{\tilde{\cal O
}}  \bar{{\cal O}}
= \bar O\bar{\chi} +\tilde\chi^\star  \Sigma
+ O\bar{\tilde\chi} +\chi^\star  \eta,
\ \frac{1}{2}\frac{\delta}{\delta\bar\theta}\frac{\delta}{\delta\theta}\bar{\cal T}\bar{\cal T}
= T\bar{\varphi} -\varphi^\star ( \phi -\chi +\bar\chi ),
\nonumber\\
&&\frac{1}{2}\frac{\delta}{\delta\bar\theta}\frac{\delta}{\delta\theta}\bar{\cal S}_
\mu  \bar{\cal S}^\mu
= S_\mu\bar{\beta}^\mu -\beta_\mu^\star  \eta^\mu. 
\end{eqnarray} 
The Lagrangian density $\bar {\cal L}_{gf}$ given in Eq. (\ref{la}) can be written using the above relations  as 
\begin{eqnarray}
\bar {\cal L}_{gf} 
=\frac{1}{2}\frac{\delta}{\delta\bar\theta}\frac{\delta}{\delta\theta}\left[
\bar {\cal B}_{\mu\nu} \bar {\cal B}^{\mu\nu} +2\bar{\tilde {\cal M}}_\mu\bar {\cal M}^\mu +\bar {\cal N}\bar {\cal N}
+\bar{\tilde {\cal N}}\bar{\tilde {\cal N}}+2\bar{\tilde{\cal O}}\bar{\cal O} +\bar{\cal T}\bar{\cal T}
+\bar{\cal S}_\mu\bar{\cal S}^\mu\right].\label{barl}
\end{eqnarray}
This implies $\bar {\cal L}_{gf} $ is  $\theta\bar\theta$ component of the superfield in the square
bracket of Eq. (\ref{barl}). Hence the $\bar {\cal L}_{gf} $ is invariant under extended BRST as well as
extended anti-BRST transformations.
Furthermore we define the super gauge-fixed fermion  as
\begin{equation}
\Gamma  (x, \theta, \bar\theta )=\Psi +\theta s_b\Psi +\bar\theta s_{ab}\Psi +\theta
\bar\theta s_b s_{ab}\Psi,
\end{equation} 
to express the ${\cal L}_{gf}$ as $\frac{\delta}{\delta\theta}\left[\Gamma  (x, \theta, 
\bar\theta) \right]$. 
The  $\theta\bar\theta$ component of $\Gamma  (x, \theta, \bar\theta )$  vanishes due to equations of motion
in the theories having both BRST and anti-BRST invariance. 
 
The complete gauge-fixed Lagrangian density  which is invariant under both extended BRST and extended anti-BRST 
transformations  can therefore be written as
\begin{eqnarray} 
{\cal L}_{gf}+\bar {\cal L}_{gf}
&=&\frac{1}{2}\frac{\delta}{\delta\bar\theta}\frac{\delta}{\delta\theta}\left[
\bar {\cal B}_{\mu\nu} \bar {\cal B}^{\mu\nu} +2\bar{\tilde {\cal M}}_\mu\bar {\cal M}^\mu +\bar {\cal N}\bar {\cal N}
+\bar{\tilde {\cal N}}\bar{\tilde {\cal N}}+2\bar{\tilde{\cal O}}\bar{\cal O} +\bar{\cal T}\bar{\cal T}
+\bar{\cal S}_\mu\bar{\cal S}^\mu\right]\nonumber\\
&+&\frac{\delta}{\delta\theta}\left[
\Gamma (x, \theta, \bar\theta)\right]. 
\end{eqnarray}
 Using equations of motion for auxiliary fields and the ghost fields associated with shift symmetry
the antifields can be calculated. With these antifields the original gauge-fixed Lagrangian density in BV 
formulation can be recovered.   
\section{\large BV formulation of 3-form gauge theory in superspace} 
The study of the higher dimensional   3-form gauge theories
is important as it appears in the excitations of
the quantized versions of strings, superstrings and related extended objects \cite{green, pol}.
In this section we will follow the same technique, as in the previous section, to develop the
superspace formulation of  extended BRST and anti-BRST invariant
Abelian 3-form gauge theory in a covariant manner. 
\subsection{\large Extended BRST invariant BV action}
Here we start with the classical Lagrangian density for the Abelian 3-form gauge theory
  as
\begin{equation}
{\cal L}_0 =\frac{1}{24} H_{\mu\nu\eta\xi}H^{\mu\nu\eta\xi},
\end{equation}
where the field strength (curvature) tensor in terms of totally antisymmetric tensor gauge 
field $B_{\mu\nu\eta}$ is defined as
\begin{equation}
H_{\mu\nu\eta\xi}= \partial_\mu B_{\nu\eta\xi} -\partial_\nu B_{ \eta\xi\mu} +\partial_\eta B_{\xi\mu\nu} 
-\partial_\xi B_{\mu\nu\eta}.
\end{equation}
This Lagrangian density 
is invariant under the infinitesimal gauge transformation for the gauge field $ B_{\mu\nu\eta}$ can be written as
\begin{equation}
\delta B_{\mu\nu\eta} =\partial_\mu \lambda_{\nu\eta} +\partial_\nu\lambda_{\eta\mu}+\partial_\eta\lambda_{\mu\nu},
\end{equation}
where $\lambda_{\mu\nu}$ is an arbitrary antisymmetric parameter.
To write the absolutely anticommuting BRST and anti-BRST invariant BV action for 3-form theory,
we consider the two equivalent candidates  for the gauge-fixing part including ghost term 
of the the Lagrangian density  
 as \cite{lm} 
\begin{eqnarray}
{\cal L}^B_{gf} 
&=&  (\partial_\mu B^{\mu\nu\eta})B_{\nu\eta} +
\frac{1}{2}B_{\mu\nu}\tilde B^{\mu\nu} +
(\partial_\mu \tilde c_{\nu\eta} +\partial_\nu \tilde c_{\eta\mu}
 + \partial_\eta \tilde c_{\mu\nu})\partial ^\mu 
c^{\nu\eta} \nonumber\\
&-&(\partial_\mu\tilde \beta_\nu -\partial_\nu \tilde\beta_\mu )\partial^\mu\beta^\nu 
 - BB_2 -\frac{1}{2} B_1^2 +(\partial_\mu \tilde c^{\mu\nu})f_\nu -(\partial_\mu c^{\mu\nu})\tilde F_\nu
\nonumber\\
&+&\partial_\mu\tilde c_2 \partial^\mu c_2 
 + \tilde f_\mu f^\mu -\tilde F_\mu F^\mu +(\partial_\mu\beta^\mu) B_2 +(\partial_\mu \phi^\mu) B_1 -(
\partial_\mu\tilde\beta^\mu) B,\\\label{lag3}
{\cal L}^{\bar B}_{gf} 
&=&  -(\partial_\mu B^{\mu\nu\eta})\bar B_{\nu\eta} +
\frac{1}{2}B_{\mu\nu}\tilde B^{\mu\nu} +
(\partial_\mu \tilde c_{\nu\eta} +\partial_\nu \tilde c_{\eta\mu}
 + \partial_\eta \tilde c_{\mu\nu})\partial ^\mu 
c^{\nu\eta} \nonumber\\
&-&(\partial_\mu\tilde \beta_\nu -\partial_\nu \tilde\beta_\mu )\partial^\mu\beta^\nu 
 - BB_2 -\frac{1}{2} B_1^2 -(\partial_\mu \tilde c^{\mu\nu})F_\nu +(\partial_\mu c^{\mu\nu})\tilde f_\nu
\nonumber\\
&+&\partial_\mu\tilde c_2 \partial^\mu c_2 
 + \tilde f_\mu f^\mu -\tilde F_\mu F^\mu +(\partial_\mu\beta^\mu) B_2 +(\partial_\mu \phi^\mu) B_1 -(
\partial_\mu\tilde\beta^\mu) B,\label{lag32}
\end{eqnarray}
where ghost field $c_{\mu\nu}$  and antighost field  $\tilde c_{\mu\nu}$ are the fermionic in nature,
however,  the vector field 
 $\phi_\mu$, antisymmetric auxiliary fields  $B_{\mu\nu}, \tilde B_{\mu\nu}$  and  auxiliary fields  $B, 
B_1, B_2$  are bosonic in nature. 
These two BV actions  are equivalent in the sense that they describe the same dynamics of the theory
on the following  CF type restricted surface
\begin{eqnarray}
 f_\mu +F_{\mu} =\partial_\mu c_1,\ \ \bar f_\mu +\bar F_{\mu} =\partial_\mu \bar c_1,\ \
 B_{\mu\nu}+\bar B_{\mu\nu} =\partial_\mu \phi_\nu -\partial_\nu\phi_\mu.
\end{eqnarray}
Both the Lagrangian densities can be written in  both BRST and anti-BRST exact terms as  
\begin{eqnarray}
{\cal L}^B_{gf}  &=& 
s_bs_{ab} \left[\frac{1}{2}\tilde c_2 c_2 -\frac{1}{2}\tilde c_1 c_1 -\frac{1}{2}\tilde c_{\mu\nu} c^{\mu\nu}-
\tilde\beta_\mu \beta^\mu -\frac{1}{2}\phi_\mu\phi^\mu -\frac{1}{6}B_{\mu\nu\eta}B^{\mu\nu\eta} \right].\\
{\cal L}^{\bar B}_{gf}  &=& 
-s_{ab}s_b \left[\frac{1}{2}\tilde c_2 c_2 -\frac{1}{2}\tilde c_1 c_1 -\frac{1}{2}\tilde c_{\mu\nu} c^{\mu\nu}-
\tilde\beta_\mu \beta^\mu -\frac{1}{2}\phi_\mu\phi^\mu -\frac{1}{6}B_{\mu\nu\eta}B^{\mu\nu\eta} \right].
\end{eqnarray}
The absolute anticommuting BRST $(s_b)$  and anti-BRST  $(s_{ab})$ transformations, which leave the   
Lagrangian densities given in Eqs. (\ref{lag3}) and (\ref{lag32}) invariant, are 
\begin{eqnarray}
s_b B_{\mu\nu\eta} &=& (\partial_\mu c_{\nu\eta}+\partial_\nu c_{\eta\mu} +\partial_\eta c_{\mu\nu}), \ \
s_b c_{\mu\nu}  =  \partial_\mu\beta_\nu -\partial_\nu \beta_\mu,
 s_b\tilde c_{\mu\nu}=B_{\mu\nu}, \nonumber\\
s_b\tilde B_{\mu\nu}  &=&\partial_\mu f_\nu -\partial_\nu f_\mu,  \ \ \ 
s_b\tilde\beta_\mu = \tilde F_\mu,\
s_b\beta_\mu =\partial_\mu  c_2,   \ \  
s_b F_\mu =-\partial_\mu B,\nonumber\\
s_b\tilde f_\mu &=& \partial_\mu B_1,  \ \ s_b\tilde c_2 =B_2,\ s_b c_1 =-B, s_b \phi_\mu =-f_\mu,\nonumber\\
 s_b \tilde c_1 &=&
 B_1,\ \
 s_b [c_2, f_\mu, \tilde F_\mu, B, B_1, B_2, B_{\mu\nu}] =0. \\
s_{ab} B_{\mu\nu\eta} &=& (\partial_\mu \tilde c_{\nu\eta}+\partial_\nu \tilde c_{\eta\mu} +\partial_\eta 
\tilde c_{\mu\nu}),  \
s_{ab}\tilde c_{\mu\nu}  =  \partial_\mu \tilde\beta_\nu -\partial_\nu \tilde\beta_\mu,  \ \ s_{ab}  c_{\mu\nu}=
\tilde B_{\mu\nu}, \nonumber\\
s_{ab}  B_{\mu\nu}  &=&\partial_\mu\tilde f_\nu -\partial_\nu\tilde f_\mu,  \ \ \ 
s_{ab} \beta_\mu =   F_\mu, \
s_{ab}\tilde\beta_\mu  = \partial_\mu\tilde  c_2,   \ \ 
s_{ab}\tilde F_\mu =-\partial_\mu B_2,\nonumber\\
s_{ab} f_\mu &=& -\partial_\mu B_1,  \ \ s_{ab}  c_2 =B,\ s_{ab} c_1 =-B_1, s_{ab} \phi_\mu = \tilde f_\mu,\nonumber\\
s_{ab} \tilde c_1 &=& - B_2,\ \ s_{ab} [\tilde c_2, \tilde f_\mu,  F_\mu, B, B_1, B_2, \tilde B_{\mu\nu}] =0. \label
{an}
\end{eqnarray}
Since ${\cal L}^B_{gf}$ is BRST invariant it can be written in terms of BRST variation of a  $\Psi$ 
as  
\begin{eqnarray}
{\cal L}^B_{gf}&=&s_b \Psi = s_b [-\frac{1}{2}\tilde c_2 B+\frac{1}{2}B_2 c_1 -\frac{1}{2}\tilde c_1 B_1 -\frac{1}{2}
(\partial_\mu\tilde \beta_\nu -\partial_\nu\tilde\beta_\mu )c^{\mu\nu}\nonumber\\
&+& \frac{1}{2}\tilde c_{\mu\nu} B^{\mu\nu} -(\partial_\mu \tilde c_2)\beta^\mu -\tilde \beta_\mu F^\mu 
-\phi_\mu \tilde f^\mu 
 - \frac{1}{3} B_{\mu\nu\eta} (\partial^\mu c^{\nu\eta} +\partial^\nu c^{\eta\mu} +\partial^\eta c^{\mu\nu} ) ],
\end{eqnarray}
where the expression for $\Psi$ is given as
\begin{eqnarray}
\Psi &=& -\frac{1}{2}\tilde c_2 B+\frac{1}{2}B_2 c_1 -\frac{1}{2}\tilde c_1 B_1 -\frac{1}{2}
(\partial_\mu\tilde \beta_\nu -\partial_\nu\tilde\beta_\mu )c^{\mu\nu} +\frac{1}{2}\tilde c_{\mu\nu} B^{\mu\nu}\nonumber\\
&-&  (\partial_\mu \tilde c_2)\beta^\mu -\tilde \beta_\mu F^\mu -\phi_\mu \tilde f^\mu
 - \frac{1}{3} B_{\mu\nu\eta} (\partial^\mu c^{\nu\eta} +\partial^\nu c^{\eta\mu} +\partial^\eta c^{\mu\nu} ).
\label{gfff}
\end{eqnarray}
The requirement for an extended Lagrangian density in BV formulation  is  that the original Lagrangian
density should be invariant under both the original BRST transformations and the
shift transformations of the original fields. Therefore, we make the shift
transformations as follows
\begin{eqnarray} 
&&B_{\mu\nu\eta} \rightarrow   B_{\mu\nu\eta}-\bar B_{\mu\nu\eta},\   c_{\mu\nu}\rightarrow  c_{\mu\nu} 
-\bar c_{\mu\nu}, \ 
\tilde c_{\mu\nu}\rightarrow   \tilde c_{\mu\nu}-\bar {\tilde c}_{\mu\nu},\ B_{\mu\nu} \rightarrow    
B_{\mu\nu} - \bar B_{\mu\nu},\nonumber\\ 
&&
\tilde B_{\mu\nu} \rightarrow \tilde B_{\mu\nu}-\bar{\tilde B}_{\mu\nu}, \ \
\beta_\mu\rightarrow  \beta_\mu - \bar\beta_\mu,\ \
\tilde \beta_\mu \rightarrow \tilde\beta_\mu-\bar{\tilde \beta}_\mu,  \ \
F_\mu\rightarrow  F_\mu - \bar{F}_\mu,\nonumber\\
&&\tilde F_\mu \rightarrow   \tilde F_\mu  -\bar{\tilde F}_\mu,\ \
f_\mu  \rightarrow   f_\mu - \bar{f}_\mu,\ \
\tilde f_\mu  \rightarrow   \tilde f_\mu  -\bar{\tilde f}_\mu,\ \
c_2\rightarrow  c_2 - \bar c_2,\nonumber\\
&&\tilde c_2  \rightarrow  \tilde c_2  -\bar{\tilde c}_2,\  
c_1\rightarrow  c_1- \bar c_1,\
\tilde c_1  \rightarrow  \tilde c_1  -\bar{\tilde c}_1,\
\phi_\mu \rightarrow   \phi_\mu- \bar \phi_\mu,\nonumber\\
&&B   \rightarrow  B  -\bar B, \ \
B_1\rightarrow  B_1- \bar B_1,\
B_2  \rightarrow   B_2  -\bar B_2,
\end{eqnarray}
where the bar fields are the shifted one corresponding to the original fields. 
The extended BRST invariant formulation of the BV action for this theory is achieved by 
choosing the following Lagrangian density:
\begin{eqnarray}
\bar{\cal L}^B_{gf}&=& {\cal L}^B_{gf} (B_{\mu\nu\eta}-\bar B_{\mu\nu\eta}, c_{\mu\nu} 
-\bar c_{\mu\nu}, \tilde c_{\mu\nu}-\bar {\tilde c}_{\mu\nu}, B_{\mu\nu} - \bar B_{\mu\nu}, 
\tilde B_{\mu\nu}-\bar{\tilde B}_{\mu\nu}, \beta_\mu - \bar\beta_\mu, \tilde\beta_\mu-\bar{\tilde \beta}_
\mu,\nonumber\\
&& F_\mu - \bar{F}_\mu, \tilde F_\mu  -\bar{\tilde F}_\mu, f_\mu - \bar{f}_\mu, \tilde f_\mu  -\bar{\tilde f}_\mu,
 c_2 - \bar c_2, \tilde c_2  -\bar{\tilde c}_2,  c_1- \bar c_1, \tilde c_1  -\bar{\tilde c}_1, \phi_\mu- \bar 
\phi_\mu,\nonumber\\
&& B  -\bar B,  B_1- \bar B_1, B_2  -\bar B_2 ).
\end{eqnarray}
Such BV action also
remains  invariant under the extended BRST transformations of fields as follows
\begin{eqnarray}
s_b \Phi (x)= \alpha (x),\ \ s_b \bar\Phi (x)&=& \alpha (x) -\beta (x),
\end{eqnarray}
where $\Phi (x)$ represents the set of all fields and $\bar\Phi (x)$  represents
the set of shifted ones.
Here $\beta (x)$ represents the original BRST transformation of collective fields $\Phi$,
whereas $\alpha (x)$ corresponds to some collective fields which generates the 
shift in fields.
 The  extended BRST transformation for such theory is given explicitly in 
Appendix A (in Eqs. (\ref{bs1}) and  (\ref{bs2})).  

We require   all the shift fields to vanish in order to obtain the  original theory.
 To achieve this goal, we choose  the 
following  
gauge-fixing part of Lagrangian density for shift symmetry as
\begin{eqnarray}
\bar{\cal L}^B_{gf} &= & l_{\mu\nu\eta}{\bar B}^{\mu\nu\eta} -B_{\mu\nu\eta}^\star (L^{\mu\nu\eta}-
\partial^\mu c^{\nu \eta}+\partial^\mu{\bar c}^{\nu\eta} - \partial^\nu c^{  \eta\mu}+
\partial^\nu{\bar c}^{\eta\mu} -\partial^\eta c^{\mu\nu }+
\partial^\eta{\bar c}^{\mu\nu} )+\bar m_{\mu\nu}{\bar c}^{\mu\nu}\nonumber\\
& +&  \tilde {c}_{\mu\nu}^\star (M^{\mu\nu}-\partial^\mu\beta^\nu 
+\partial^\mu \bar \beta^\nu +\partial^\nu\beta^\mu -
\partial^\nu \bar \beta^\mu) + m_{\mu\nu}\bar {\tilde c}^{\mu\nu} +{ c}_{\mu\nu}^\star (\tilde M^{\mu\nu}
- B^{\mu\nu} +\bar B^{\mu\nu})\nonumber\\
&+&n_{\mu\nu}\bar B^{\mu\nu} -B_{\mu\nu}^\star N^{\mu\nu} +\bar n_{\mu\nu}\bar {\tilde B}^{\mu\nu} -
\tilde B_{\mu\nu}^\star (\tilde N^{\mu\nu} -\partial^\mu f^\nu +\partial^\mu\bar f^\nu  +\partial^\nu f^\mu -
\partial^\nu\bar f^\mu )\nonumber\\
&+& o_\mu \bar \beta ^\mu -\beta_\mu^\star (O^\mu -\partial^\mu c_2 +\partial^\mu \bar c_2 ) +\bar o_\mu \bar {\tilde 
\beta}^\mu -\tilde \beta_\mu^\star (\tilde O^\mu -\tilde F ^\mu +\bar {\tilde F}^\mu) 
+\bar p_\mu \bar F^\mu\nonumber\\
&+& \tilde F_\mu^\star (P^\mu +\partial^\mu B -\partial^\mu \bar B) + p_\mu \bar {\tilde F} ^\mu 
+  F_\mu^\star \tilde P^\mu + \bar q_\mu \bar f^\mu + \tilde f_\mu^\star Q^\mu +  q_\mu \bar {\tilde f^\mu}\nonumber\\
&+&   f_\mu^\star (\tilde Q^\mu -\partial ^\mu B_1 +\partial ^\mu \bar B_1 )+\bar r\bar c_2 +\tilde c_2^\star R 
+ r \bar {\tilde c}_2 + c_2^\star (\tilde R -B_2 +\bar B_2 ) + \bar s \bar c_1 \nonumber\\
&+& \tilde c_1^\star (S+B-\bar B) +  s \bar {\tilde c}_1 +   c_1^\star (\tilde S -B_1 +\bar B_1)+ t_\mu \bar 
\phi^\mu -\phi_\mu ^\star (T^\mu -f^\mu +\bar f^\mu ) + u\bar B \nonumber\\
&-& B^\star U +v\bar B_1 -B_1^\star V +w\bar B_2 -B_2 ^\star W,  
\end{eqnarray}
where the fields $L_{\mu\nu\eta},$  $M_{\mu\nu},$ $\tilde M_{\mu\nu},$ $N_{\mu\nu},$ $\tilde N_{\mu\nu},$ $O_\mu,$ 
$\tilde O_\mu,$ $P_\mu,$ $\tilde P_\mu,$ $Q_\mu,$ $\tilde Q_\mu,$ $R,$ $\tilde R,$ $S,$ $\tilde S,$ $T_\mu,$ $U,$ 
$V,$ $W$ are ghost fields 
associated with the shift symmetries for fields 
$B_{\mu\nu\eta},$ $c_{\mu\nu},$ $\tilde c_{\mu\nu},$ $B_{\mu\nu},$ $\tilde B_{\mu\nu},$ $\beta_\mu,$ $\tilde \beta_\mu,$ $F_\mu,$ 
$\tilde F_\mu,$ $f_\mu,$ $\tilde f_\mu,$ $c_2,$ $\tilde c_2,$ $c_1,$ $\tilde c_1,$ $\phi_\mu,$ $B,$ $B_1,$ $B_2$ 
respectively and  fields $l_{\mu\nu\eta},$  $m_{\mu\nu},$ $\bar m_{\mu\nu},$ $n_{\mu\nu},$ $\bar 
n_{\mu\nu},$ $o_\mu,$ $\bar
o_\mu,$ $p_\mu,$ $\bar p_\mu,$ $q_\mu,$  $\bar q_\mu,$ $r,$ $\bar r,$ $s,$ $\bar s,$ $t_\mu,$ $u,$ $v,$ $w$  
are Nakanishi-Lautrup type auxiliary fields corresponding to the antighost fields $B_{\mu\nu\eta}^\star,$ 
$c_{\mu\nu}^\star,$  $\tilde 
c_{\mu\nu}^\star,$ $B_{\mu\nu}^\star,$ 
$\tilde B_{\mu\nu}^\star,$ $\eta_\mu^\star,$   $\tilde \beta_\mu^\star,$  $F_\mu^\star,$ 
$\tilde F_\mu^\star,$ $f_\mu^\star,$ $\tilde f_\mu^\star,$ $c_2^\star,$ $\tilde c_2^\star,$ $c_1^\star,$ 
$\tilde c_1^\star,$ 
$\phi_\mu^\star,$ $B^\star,$ $B_1^\star,$ $B_2^\star$ respectively.

Such a Lagrangian density $\bar{\cal L}^B_{gf}$ is invariant 
under the extended BRST symmetry transformations given in Eqs. (\ref{bs1}) and (\ref{bs2}) (see Appendix).
All  bar fields disappear when we use the equations of motion for auxiliary fields and  
the extended Lagrangian density takes the form
\begin{eqnarray}
\bar{\cal L}^B_{gf} &= &  -B_{\mu\nu\eta}^\star (L^{\mu\nu\eta}-
\partial^\mu c^{\nu \eta} - \partial^\nu c^{  \eta\mu} -\partial^\eta c^{\mu\nu }  )
  +   \tilde {c}_{\mu\nu}^\star (M^{\mu\nu}-\partial^\mu\beta^\nu  +\partial^\nu\beta^\mu ) \nonumber\\
 & +& {c}_{\mu\nu}^\star (\tilde M^{\mu\nu}- B^{\mu\nu} )  -B_{\mu\nu}^\star N^{\mu\nu}  - 
\tilde B_{\mu\nu}^\star (\tilde N^{\mu\nu} -\partial^\mu f^\nu   +\partial^\nu f^\mu  )\nonumber\\
&-& \beta_\mu^\star (O^\mu -\partial^\mu c_2 ) -\tilde \beta_\mu^\star (\tilde O^\mu -\tilde F ^\mu )  
 + \tilde F_\mu^\star (P^\mu +\partial^\mu B ) 
+  F_\mu^\star \tilde P^\mu  +\tilde  f_\mu^\star Q^\mu \nonumber\\
&+& f_\mu^\star (\tilde Q^\mu -\partial ^\mu B_1  )  +\tilde c_2^\star R + 
 c_2^\star (\tilde R -B_2 ) 
 + \tilde c_1^\star (S+B) + c_1^\star (\tilde S -B_1 )\nonumber\\
&-&\phi_\mu ^\star (T^\mu -f^\mu )  
 -  B^\star U   -B_1^\star V   -B_2 ^\star W.\label{ki}
\end{eqnarray}
On the other hand, if   $\Psi$ depends on the original fields only, then a
general gauge-fixing Lagrangian density 
of Abelian rank-3 antisymmetric tensor field with the original BRST symmetry  have the following form
\begin{equation}
{\cal L}^B_{gf}  =  s_b \Psi [\Phi]= \Sigma  (s_b \Phi) \frac{\delta\Psi}{\delta \Phi},
\end{equation}
where $\Phi$ is the generic notation for all fields in the theory.
Keeping the fermionic/bosonic nature of  fields in mind the above gauge-fixed
Lagrangian density can be re-expressed as 
\begin{eqnarray}
{\cal L}^B_{gf} &= & -\frac{\delta\Psi}{\delta B_{\mu\nu\eta}}L_{\mu\nu\eta}+
 \frac{\delta\Psi}{\delta c_{\mu\nu}}M_{\mu\nu} + \frac{\delta\Psi}
{\delta\tilde{c}_{\mu\nu}}\tilde M_{\mu\nu} -
 \frac{\delta\Psi}{\delta B_{\mu\nu}}N_{\mu\nu}-\frac{\delta\Psi}{\delta \tilde B_{\mu\nu}}\tilde N_{\mu\nu}
\nonumber\\
&-& \frac{\delta\Psi}{\delta
\beta_\mu}O_\mu - \frac{\delta\Psi}{\delta
\tilde\beta_\mu}\tilde O_\mu +\frac{\delta\Psi}{\delta F_\mu}P_\mu +\frac{\delta\Psi}{\delta \tilde F_\mu}
\tilde P_\mu + \frac{\delta\Psi}{\delta f_\mu}Q_\mu +\frac{\delta\Psi}{\delta \tilde f_\mu}\tilde Q_\mu \nonumber\\
&+&
 \frac{\delta\Psi}{ \delta c_2}R + \frac{\delta\Psi}{ \delta \tilde c_2}\tilde R
 + \frac{\delta\Psi}{ \delta c_1}S + \frac{\delta\Psi}{ \delta \tilde c_1}\tilde S 
 -  \frac{\delta\Psi}{ \delta \phi_\mu}T_\mu - \frac{\delta\Psi}{ \delta B}U
-\frac{\delta\Psi}{\delta B_1}V-\frac{\delta\Psi}{\delta B_2}W. \label{ci} 
\end{eqnarray}
Combining all the  Lagrangian densities of such theory
given in Eqs. (\ref{lag3}), (\ref{ki}) and (\ref{ci}),  
the total Lagrangian density for Abelian 3-form gauge theory in BV formulation can be written as  
\begin{eqnarray}  
 {\cal L}_T&=& {\cal L}_0+ 
{\cal L}^B_{gf}+\bar{\cal L}^B_{gf}, \nonumber\\
&=& \frac{1}{12}H_{\mu \nu \eta\xi}H^{\mu \nu \eta\xi} +B_{\mu\nu\eta}^\star (
\partial^\mu c^{\nu \eta} + \partial^\nu c^{  \eta\mu} +
\partial^\eta c^{\mu\nu }  )
  -  \tilde {c}_{\mu\nu}^\star ( \partial^\mu\beta^\nu  -\partial^\nu\beta^\mu ) \nonumber\\
 & -& {c}_{\mu\nu}^\star 
 B^{\mu\nu}   +\tilde B_{\mu\nu}^\star (\partial^\mu f^\nu   -\partial^\nu f^\mu  ) 
+\beta_\mu^\star \partial^\mu c_2  +\tilde \beta_\mu^\star \tilde F ^\mu 
 + \tilde F_\mu^\star  \partial^\mu B  - f_\mu^\star  \partial ^\mu B_1   \nonumber\\ 
&-&  
 c_2^\star   B_2  
 +\tilde c_1^\star  B  - c_1^\star  B_1 
+\phi_\mu ^\star   f^\mu    
 - \left(B^{\mu\nu\eta\star}+\frac{\delta\Psi}{\delta B_{\mu\nu\eta}}\right)L_{\mu\nu\eta}+
\left( \tilde c^{\mu\nu\star} +
 \frac{\delta\Psi}{\delta c_{\mu\nu}}\right) M_{\mu\nu} \nonumber\\
&+&\left( c^{\mu\nu\star} + \frac{\delta\Psi}
{\delta\tilde{c}_{\mu\nu}}\right)\tilde M_{\mu\nu} -\left(B^{\mu\nu\star} +
 \frac{\delta\Psi}{\delta B_{\mu\nu}}\right) N_{\mu\nu}
-\left(\tilde B^{\mu\nu\star} +\frac{\delta\Psi}{\delta \tilde B_{\mu\nu}}\right) 
\tilde N_{\mu\nu}\nonumber\\
&-&\left(\beta^{\mu\star} +\frac{\delta\Psi}{\delta
\beta_\mu}\right)O_\mu - \left(\tilde\beta^{\mu\nu\star} +\frac{\delta\Psi}{\delta
\tilde\beta_\mu}\right)\tilde O_\mu +\left( \tilde F^{\mu\star} +\frac{\delta\Psi}{\delta F_\mu}\right)P_\mu 
\nonumber\\
&+&\left( F^{\mu\star} +\frac{\delta\Psi}{\delta \tilde F_\mu}\right)
\tilde P_\mu  
 +  \left( \tilde f^{\mu \star} +\frac{\delta\Psi}{\delta f_\mu}\right)Q_\mu +\left( f^{\mu\star} 
+\frac{\delta\Psi}{\delta \tilde f_\mu}\right)\tilde Q_\mu
 + \left( \tilde c_2^{\star} +\frac{\delta\Psi}{ \delta c_2}\right) R \nonumber\\
&+&\left( c_2^{\star} + \frac{\delta\Psi}{ \delta \tilde c_2}\right)\tilde R
 + \left( \tilde c_1^{\star} +\frac{\delta\Psi}{ \delta c_1}\right)S 
+ \left(c_1^{\star} +\frac{\delta\Psi}{ \delta \tilde c_1}\right)\tilde S-
 \left(\phi^{\mu\star} +\frac{\delta\Psi}{ \delta \phi_\mu}\right) T_\mu\nonumber\\
 &-&  \left( B^{\star} +\frac{\delta\Psi}{ \delta B}\right) U
-\left(B_1^{\star} +\frac{\delta\Psi}{\delta B_1}\right) V-\left(B_2^{\star} +\frac{\delta\Psi}{\delta B_2}
\right) W.
\end{eqnarray}
The antighost fields are identified as the antifields after eliminating 
ghost fields associated with the shift symmetry and are given as
\begin{eqnarray}
&&B^{\mu\nu\eta\star}=\frac{1}{3} (\partial^\mu c^{\nu\eta}
+\partial^\nu c^{\eta\mu} +\partial^\eta c^{\mu\nu} ),\   c^{\mu\nu\star}  = -\frac{1}{2}B^{\mu\nu},
\ \ B^{\mu\nu\star} =-\frac{1}{2}\tilde c^{\mu\nu},\nonumber\\
&&\tilde c^{\mu\nu\star} =\frac{1}{2}(\partial^\mu\tilde \beta^\nu -\partial^\nu\tilde\beta^\mu )-
\partial_\eta B^{\mu\nu\eta},\ \
\beta^{\mu\star} =-\frac{1}{2}\partial^\mu\tilde c_2,\ \
\tilde\beta^{\mu \star} =F^\mu +\partial_\nu c^{\mu\nu},\nonumber\\  
&&\tilde F^{\mu\star} =\tilde\beta^\mu,\ 
 f^{\mu\star}  = \phi^\mu,\ 
c_2^{\star}=\frac{1}{2}B-\partial_\mu\beta^\mu,
\  \tilde c_1^{\star} =-\frac{1}{2}B_2,\ \
 c_1^{\star} =\frac{1}{2}B_1,\nonumber\\
&&\phi^{\mu\star} =\tilde f^\mu,\
B^{\star} =\frac{1}{2}\tilde c_2,\ 
  B_1^{\star} =\frac{1}{2}\tilde c_1,\
B_2^{\star} = -\frac{1}{2}c_1,\ [\tilde B^{\mu\nu\star}, F^{\mu\star}, f^{\mu \star}, \tilde c_2^{\star}]=0 .
\end{eqnarray}
We are now able to write the gauge-fixed part of total Lagrangian density as a BRST
variation of a generalized gauge-fixing fermion 
\begin{eqnarray}
 {\cal L}^B_{gf} +\bar{\cal L}^B_{gf}&=&  s_b \left(B_{\mu\nu\eta}^\star \bar B^{\mu\nu\eta} +
c_{\mu\nu}^\star \bar {\tilde c}^{\mu\nu} +\tilde c_{\mu\nu}^\star \bar {c}^{\mu\nu} +
B_{\mu\nu}^\star \bar B^{\mu\nu} +\beta_{\mu}^\star \bar \beta^{\mu} +\tilde\beta_{\mu}^\star 
\bar{\tilde \beta}^{\mu}+F_{\mu}^\star \bar{\tilde F}^{\mu}+ \tilde F_{\mu}^\star \bar{ F}^{\mu}\right.\nonumber\\
&+&\left.f_{\mu}^\star \bar{\tilde f}^{\mu}+ \tilde f_{\mu}^\star \bar{ f}^{\mu} 
 +
c_2^\star \bar{\tilde c_2} + \tilde c_2^\star \bar{ c_2} 
 + c_1^\star \bar{\tilde c_1} + \tilde c_1^\star \bar{ c_1} 
+\phi_\mu^\star\bar\phi^\mu +B^\star\bar B+B_1^\star\bar B_1+B_2^\star\bar B_2
\right),\nonumber\\ 
&=&s_b \left(\Phi^\star\bar\Phi\right),
\end{eqnarray}
where the fields $\Phi$ and $\bar\Phi$ are the generic notation 
for all original fields and corresponding shifted fields respectively,
the  ghost number  of the expression  $\Phi^\star\bar\Phi=-1$  as expected. Here we note the 
difference  with the ordinary gauge-fixing fermion given as
\begin{eqnarray}
\Psi &=& - [ B_{\mu\nu\eta}B^{\mu\nu\eta\star} +\tilde c_{\mu\nu} c^{\mu\nu\star} +\tilde\beta_\mu
\tilde\beta^{\mu\star}+\phi_\mu\phi^{\mu\star}+\tilde c_2c_2^\star +\tilde c_1 c_1^\star +B_2 B_2^\star].
\end{eqnarray} 
\subsection{\large Superspace formulation:  Extended BRST invariant BV action } 
In this subsection we develop a superspace
formalism for extended BRST invariant BV action developed in the previous subsection. For this purpose we  
consider  superspace with one 
fermionic parameter $\theta$ and
define the following superfields, $\Upsilon $,  in terms of generic fields $\Phi$
\begin{eqnarray}
\Upsilon  (x, \theta) =\Phi (x) +\theta (s_b\Phi ).\label{brss}
\end{eqnarray}
Explicit expressions for each superfield are listed in Appendix (see Eq. (\ref{supb})). 

The gauge-fixing part of the
Lagrangian density given in Eq. (\ref{lag3}) is written, in this 
superspace formulation as
\begin{eqnarray}
\bar{\cal L}^B_{gf} &=& \frac{\delta}{\delta\theta}\left[\bar{\cal B}_{\mu\nu\eta}^\star \bar{\cal 
B}^{\mu\nu\eta} +\bar{\tilde{\cal C}}_{\mu\nu}^\star  \bar{\cal C}^{\mu\nu}
+\bar{\tilde{\cal C}}_{\mu\nu}  \bar{\cal C}^{\mu\nu\star}+\bar{\tilde{\cal B}}_{\mu\nu}^\star  
\bar{\tilde{\cal B}}^{\mu\nu}
+\bar{{\cal B}}_{\mu}^\star  \bar{\cal B}^{\mu}+ \bar{\tilde{\cal F}}_\mu^\star \bar {\cal F}^\mu +\bar{\tilde{\bf f}}_\mu^\star \bar {\bf f}^\mu 
\right.\nonumber\\
&+&\left. \bar{\tilde{\cal F}}_\mu  \bar {\cal F}^{\mu\star} +\bar{\tilde{\bf f}}_\mu \bar {\bf f}^{\mu\star}
 +\bar{\tilde{\cal C}}_1^\star  \bar{\cal 
C}_1  + \bar{\tilde{\cal C}}_1 \bar{\cal 
C}^\star _1+ \bar{\tilde{\cal C}}_2^\star  \bar{\cal 
C}_2  + \bar{\tilde{\cal C}}_2 \bar{\cal 
C}^\star _2+\bar{\cal B}^\star  \bar{\cal 
B} +\bar{\cal B}_1^\star  \bar{\cal 
B}_1  +\bar{\cal B}_2^\star  \bar{\cal 
B}_2\right].
\end{eqnarray}
$\bar{\cal L}^B_{gf}$ 
remains invariant under the
extended BRST transformation as it is $\theta$ component of a superfield. The gauge-fixed Lagrangian density for the
original symmetry can also be written in this formalism by defining  $\Gamma $ as
\begin{equation}
{\Gamma  } =\Psi +\theta s_b \Psi. 
\end{equation}
Assuming $\Psi$ is a function of all original fields, we write
\begin{eqnarray}
\Gamma  &=&\Psi +\theta \left[-\frac{\delta\Psi}{\delta B_{\mu\nu\eta}}L_{\mu\nu\eta}+
 \frac{\delta\Psi}{\delta c_{\mu\nu}}M_{\mu\nu} + \frac{\delta\Psi}
{\delta\tilde{c}_{\mu\nu}}\tilde M_{\mu\nu} -
 \frac{\delta\Psi}{\delta B_{\mu\nu}}N_{\mu\nu}
-\frac{\delta\Psi}{\delta \tilde B_{\mu\nu}}\tilde N_{\mu\nu}\right.\nonumber\\
&-&\left. \frac{\delta\Psi}{\delta
\beta_\mu}O_\mu - \frac{\delta\Psi}{\delta
\tilde\beta_\mu}\tilde O_\mu +\frac{\delta\Psi}{\delta F_\mu}P_\mu +\frac{\delta\Psi}{\delta \tilde F_\mu}
\tilde P_\mu + \frac{\delta\Psi}{\delta f_\mu}Q_\mu +\frac{\delta\Psi}{\delta \tilde f_\mu}\tilde Q_\mu
 + \frac{\delta\Psi}{ \delta c_2}R
\right.\nonumber\\
&+&  \left. \frac{\delta\Psi}{ \delta \tilde c_2}\tilde R
 + \frac{\delta\Psi}{ \delta c_1}S + \frac{\delta\Psi}{ \delta \tilde c_1}\tilde S-
 \frac{\delta\Psi}{ \delta \phi_\mu}T_\mu - \frac{\delta\Psi}{ \delta B}U
-\frac{\delta\Psi}{\delta B_1}V-\frac{\delta\Psi}{\delta B_2}W \right].
\end{eqnarray}
Thus, we write the original gauge-fixing Lagrangian density in the superspace
formalism as
\begin{equation}
{\cal L}^B_{gf} =\frac{\delta\Gamma }{\delta\theta}.
\end{equation}
Once again, being the $\theta$ component of a superfield, this is manifestly invariant
under the extended BRST transformation. Now, we are able to write the total Lagrangian density
in the superspace as 
\begin{eqnarray}
\bar{\cal L}^B_{gf} & =&{\cal L}_0+\bar{\cal L}^B_{gf} +{\cal L}^B_{gf} \nonumber\\
& =&{\cal L}_0 + \frac{\delta}{\delta\theta}\left[\bar\Upsilon ^\star \bar\Upsilon  \right]
+\frac{\delta\Gamma }{\delta\theta},                                        
\end{eqnarray} 
where $ \bar\Upsilon ^\star$ and $\bar\Upsilon $ are the generic notation for the shift fields 
corresponding to the super antifields $ \Upsilon^\star$ and superfields $\Upsilon $ respectively. 
This Lagrangian density is manifestly invariant under the original BRST symmetry, after
elimination of the auxiliary and ghost fields associated with the shift symmetry.
\subsection{\large  Extended anti-BRST invariant BV action} 
In the previous subsections we have analyzed the extended BRST symmetry for the Lagrangian
density of the Abelian rank-3 antisymmetric tensor field
and corresponding superspace formulation. In this subsection we study the extended 
anti-BRST symmetry for such theory.

The gauge-fixed anti-fermion $\bar\Psi$ for the theory is defined as
\begin{eqnarray}
\bar\Psi &=& -\frac{1}{2} B_2 c_2 +\frac{1}{2} B_1c_1 +\frac{1}{2} \tilde c_1 B +\frac{1}{2} B_{\mu\nu} c^{\mu\nu} -
\frac{1}{2}\tilde c_{\mu\nu} (\partial^\mu\beta^\nu -\partial^\nu \beta^\mu ) +\tilde F_\mu \beta^\mu +\beta _\mu 
\partial^\mu c_2\nonumber\\
& +&\frac{1}{2} \phi_\mu f^\mu +\frac{1}{3} B_{\mu\nu\eta} (\partial^\mu c^{\nu\eta} +\partial^\nu c^{
\eta\mu} +\partial^\eta c^{\mu\nu} ).
\end{eqnarray} 
So that the gauge-fixing part of the Lagrangian density can be
written in terms of $\bar\Psi$ as
\begin{eqnarray}
{\cal L}^{\bar B}_{gf} = s_{ab} \bar\Psi &= &
  s_{ab} \left[ -\frac{1}{2} B_2 c_2 +\frac{1}{2} B_1c_1 +\frac{1}{2} \tilde c_1 B +\frac
{1}{2} B_{\mu\nu} c^{\mu\nu} -
\frac{1}{2}\tilde c_{\mu\nu} (\partial^\mu\beta^\nu -\partial^\nu \beta^\mu ) +\tilde F_\mu \beta^\mu 
\right.\nonumber\\
& +&\left.\beta _\mu 
\partial^\mu c_2+\frac{1}{2} \phi_\mu f^\mu +\frac{1}{3} B_{\mu\nu\eta} (\partial^\mu c^{\nu\eta} +\partial^\nu c^{
\eta\mu} +\partial^\eta c^{\mu\nu} )\right].
\end{eqnarray}
Following the  structure of Eq. (\ref{bs1}) given in Appendix A, we demand that $s_{ab}(\Phi-\bar\Phi)$ 
reproduces the anti-BRST transformations of ordinary fields $(\Phi)$ for
the Abelian rank-3 antisymmetric tensor field theory mentioned in Eq. (\ref{an}), 
consequently we 
 have following transformations  
\begin{eqnarray}
&&s_{ab} \bar {B}_{\mu\nu\eta} = B_{\mu\nu\eta}^\star,\ s_{ab} {B}_{\mu\nu\eta}= B_{\mu\nu\eta}^\star +
\partial_\mu \tilde c_{\nu\eta}-\partial_\mu \bar{\tilde c}_{\nu\eta}+\partial_\nu \tilde c_{\eta\mu}
-\partial_\nu \bar{\tilde c}_{\eta\mu}  +\partial_\eta 
\tilde c_{\mu\nu}-\partial_\eta \bar{\tilde c}_{\mu\nu},\nonumber\\
 &&s_{ab} \bar{\tilde c}_{\mu\nu} = \tilde c_{\mu\nu}^\star,\ \
s_{ab} {\tilde c}_{\mu\nu} =\tilde c_{\mu\nu}^\star +\partial_\mu \tilde\beta_\nu -\partial_\mu 
\bar{\tilde\beta}_\nu -\partial_\nu \tilde\beta_\mu +\partial_\nu \bar{\tilde\beta}_\mu,\ \
s_{ab} \bar{ c}_{\mu\nu} =c_{\mu\nu}^\star,\nonumber\\
&&s_{ab} { c}_{\mu\nu} = c_{\mu\nu}^\star +\tilde B_{\mu\nu} -\bar{\tilde B}_{\mu\nu}, \ \
s_{ab} { B}_{\mu\nu} = B_{\mu\nu}^\star + \partial_\mu\tilde f_\nu -\partial_\mu\bar{\tilde f}_\nu
-\partial_\nu\tilde f_\mu +\partial_\nu\bar{\tilde f}_\mu,\nonumber\\
 &&s_{ab} \bar{ B}_{\mu\nu} = B_{\mu\nu}^\star,\ \
s_{ab} \bar \beta_\mu =\beta_\mu^\star,\ \ s_{ab} \beta_\mu
=\beta_\mu^\star +F_\mu -\bar F_\mu, \ \
s_{ab} \bar {\tilde \beta}_\mu  = \tilde\beta_\mu^\star,\nonumber\\
&& s_{ab} \tilde \beta_\mu =\tilde\beta_\mu^\star +\partial_\mu \tilde c_2-\partial_\mu \bar{\tilde c}_2,
\ \ s_{ab} \bar {\tilde F}_\mu  = \tilde F_\mu^\star,\ \ s_{ab} \tilde F_\mu
=\tilde F_\mu^\star -\partial_\mu B_2+\partial_\mu \bar B_2,\nonumber\\
 && s_{ab} \bar {f}_\mu  =f_\mu^\star,\ \ s_{ab} f_\mu
=f_\mu^\star -\partial_\mu B_1+\partial_\mu \bar B_1,\ \
s_{ab} \bar {c}_2 =c_2^\star,\ \ s_{ab} c_2
=c_2^\star +B -\bar B,\nonumber\\ 
&& s_{ab} \bar {c}_1  =c_1^\star,\ \
s_{ab} c_1 =c_1^\star -B_1 +\bar B_1,\ \ s_{ab} \bar {\phi}_\mu  = 
\phi_\mu^\star,\ \ s_{ab} \phi_\mu  = \phi_\mu^\star +\tilde f_\mu -\bar{\tilde f}_\mu,\nonumber\\
&&s_{ab} \bar {\tilde c}_1  =\tilde c_1^\star,\ \ s_{ab} \tilde c_1
=\tilde c_1^\star -B_2 +\bar B_2,\ \ 
s_{ab} \bar {\tilde c}_2 =\tilde c_2^\star,\ \ s_{ab} \tilde c_2
=\tilde c_2^\star, \ \
s_{ab} \bar {\tilde f}_\mu  = \tilde f_\mu^\star,\nonumber\\
 &&s_{ab} \tilde f_\mu
=\tilde f_\mu^\star,\ \ 
 s_{ab} \bar {F}_\mu   = F_\mu^\star,\   s_{ab} F_\mu
=F_\mu^\star, \  s_{ab} \bar B  = B^\star,\ \
s_{ab} B =B^\star, \ 
 s_{ab} \bar B_1  = B_1^\star,\nonumber\\
&& s_{ab} B_1
=B_1^\star,\ 
 s_{ab} \bar B_2 =B_2^\star,\  s_{ab} B_2
=B_2^\star,\ 
s_{ab} \bar {\tilde B}_{\mu\nu}  = \tilde B_{\mu\nu}^\star,    s_{ab} \tilde B_{\mu\nu}
=\tilde B_{\mu\nu}^\star,\
 s_{ab}\Xi =0,\nonumber\\ 
 &&\Xi \equiv [ B_{\mu\nu\eta}^\star, \tilde c_{\mu\nu}^\star,  c_{\mu\nu}^\star, B_{\mu\nu}^\star,
 \beta_\mu^\star, \tilde \beta_\mu^\star, \tilde F_\mu^\star, f_\mu^\star, c_2^\star, c_1^\star,
 \phi_\mu^\star, \tilde c_1^\star, \tilde c_2^\star, \tilde f_\mu^\star, F_\mu^\star, B^\star, B_1^\star, B_2^\star,
 \tilde B_{\mu\nu}^\star ]. 
\end{eqnarray}
The ghost fields associated with the shift symmetry have the following extended
anti-BRST transformations,
\begin{eqnarray}
s_{ab}L_{\mu\nu\eta}&=& l_{\mu\nu\eta},\ \ s_{ab}M_{\mu\nu} = m_{\mu\nu},\ \
 s_{ab}\tilde M_{\mu\nu} =\bar m_{\mu\nu},\ \
s_{ab}N_{\mu\nu} = n_{\mu\nu},\nonumber\\
 s_{ab}\tilde N_{\mu\nu} &=&\bar n_{\mu\nu},\ \
s_{ab}O_{\mu} = o_{\mu},\  
s_{ab}\tilde O_{\mu} = \bar o_{\mu},\  
 s_{ab}P_{\mu} =p_{\mu},\
s_{ab}\tilde P_{\mu} = \bar p_{\mu},\nonumber\\
 s_{ab}Q_{\mu} &=&q_{\mu},\ \
s_{ab}\tilde Q_{\mu} = \bar q_{\mu},\ \ 
s_{ab}R   = r,\ \
 s_{ab}\tilde R =\bar r,\ \
s_{ab}S = s,\nonumber\\
s_{ab}\tilde S  &=& \bar s,\ \
s_{ab}T_\mu =t_\mu,\ \
s_{ab}U = u,\ \ 
s_{ab}V   = v,\ \ s_{ab}W   = w.
\end{eqnarray}
 
\subsection{\large Extended BRST and anti-BRST invariant superspace formulation}
To write a Lagrangian density that is manifestly invariant under both extended
BRST transformations and  extended anti-BRST transformations
in superspace formalism  we introduce a superspace with   two Grassmann
parameters,   $\theta$ and $\bar\theta$. The generic superfields in this superspace are defined as
\begin{equation}
\Upsilon (x, \theta, \bar\theta )=\Phi (x) +\theta (s_b \Phi) +\bar\theta (s_{ab} \Phi)
+\theta\bar\theta (s_bs_{ab}\Phi ),
\end{equation} 
where $\Upsilon $ and $\Phi$ are the generic notation for all the superfields and the fields respectively.
The expressions of all the individual superfields are given in the Appendix A (see Eq. \ref{superf}).   

One can write the gauge-fixing Lagrangian density $\bar {\cal L}^B_{gf}$ 
given in Eq. (\ref{lag3}) in terms of these superfields as 
\begin{eqnarray}
\bar {\cal L}^B_{gf} 
&=&\frac{1}{2}\frac{\delta}{\delta\bar\theta}\frac{\delta}{\delta\theta}\left[
\bar {\cal B}_{\mu\nu\eta} \bar {\cal B}^{\mu\nu\eta} +2\bar{\tilde {\cal C}}_{\mu\nu}\bar {\cal C}^{\mu\nu} 
+ \bar{\tilde {\cal B}}_{\mu\nu} \bar{\tilde {\cal B}}^{\mu\nu}+\bar{\cal B}_{\mu\nu} \bar{\cal B}^{\mu\nu}
 +\bar{\cal B}_{\mu } \bar{\cal B}^{\mu}+\bar{\tilde {\cal B}}_{\mu } 
\bar{\tilde{\cal B}}^{\mu}\right.\nonumber\\
&+&\left. 2\bar{\tilde {\cal F}}_{\mu } \bar{\cal F}^{\mu}+ 2\bar{\tilde {\bf f}}_{\mu } \bar{\bf f}^{\mu} 
 + 2\bar{\tilde {\cal C}}_{2 } \bar
{\cal C}_2 +2\bar{\tilde {\cal C}}_{1} \bar{{\cal C}}_1+\bar{\Phi}_{\mu } \bar{ \Phi}^{\mu}+\bar {\cal B}\bar {\cal B}
+\bar {\cal B}_1\bar {\cal B}_1 +\bar {\cal 
B}_2\bar {\cal B}_2\right],
\end{eqnarray}
This implies that the Lagrangian density $\bar {\cal L}^B_{gf} $
is manifestly invariant under extended BRST and anti-BRST transformations.
Furthermore, we   define the gauge-fixing fermion in superspace as
\begin{equation}
\Gamma  (x, \theta, \bar\theta )=\Psi +\theta s_b\Psi +\bar\theta s_{ab}\Psi +\theta
\bar\theta s_b s_{ab}\Psi.
\end{equation} 
The component of $\theta\bar\theta$ disappears from the above expression using equations of motion
and therefore the Lagrangian
density for the original fields can be written as
\begin{equation}
{\cal L}^B_{gf} =\frac{\delta}{\delta\theta}\left[\Gamma  (x, \theta, 
\bar\theta) \right] = s_b\Psi.
\end{equation}
This Lagrangian density is not only manifestly invariant under extended BRST
transformations but also invariant under extended anti-BRST transformations. 
The complete gauge-fixed Lagrangian density can therefore be written as
\begin{eqnarray}
{\cal L}^B_{gf}+\bar {\cal L}^B_{gf}
&=&\frac{1}{2}\frac{\delta}{\delta\bar\theta}\frac{\delta}{\delta\theta}\left[
\bar {\cal B}_{\mu\nu\eta} \bar {\cal B}^{\mu\nu\eta} +2\bar{\tilde {\cal C}}_{\mu\nu}\bar {\cal C}^{\mu\nu} 
+ \bar{\tilde B}_{\mu\nu} \bar{\tilde B}^{\mu\nu}+\bar{ B}_{\mu\nu} \bar{ B}^{\mu\nu}
 +\bar{\beta}_{\mu } \bar{\beta}^{\mu}+\bar{\tilde \beta}_{\mu } \bar{\tilde\beta}^{\mu}\right.\nonumber\\
&+&\left. 2\bar{\tilde F}_{\mu } \bar{F}^{\mu}   + 2\bar{\tilde f}_{\mu } \bar{f}^{\mu} + 2\bar{\tilde c}_{2 } \bar
{c}_2 +2 \bar{\tilde c}_{1} \bar{c}_1 +\bar{\phi}_{\mu } \bar{\phi}^{\mu}+\bar B\bar B+\bar B_1\bar B_1 +\bar B_2\bar 
B_2\right]\nonumber\\
&+&\frac{\delta}{\delta\theta}\left[
\Gamma (x, \theta, \bar\theta)\right].
\end{eqnarray}
Using equations of motion of auxiliary fields the tilde fields can made vanish and
by integrating out the ghost fields for the shift symmetry we will get the explicit
expressions for the antifields.

\section{\large Concluding Remarks}
Higher form gauge theories play a very important role in certain string theoretic and supergravity models. 
In this work we have considered the BV formulation of extended BRST and anti-BRST 
invariant (including some shift symmetry) 2-form and 3-form gauge theories. Antifields arise naturally in such 
formulation. We have further
constructed a superspace formulation for these theories. We have shown that the BV action for 2-form
as well as 3-form gauge theories can be written in a manifestly extended BRST invariant manner in a superspace
with one fermionic coordinate. 
 However, a superspace with two Grassmann coordinates are required for a manifestly covariant formulation 
of the extended BRST and extended anti-BRST invariant BV actions for higher form gauge theories.
It will be interesting to extend this formulation for anomalous gauge theories. 

\section*{Acknowledgments}

We thankfully acknowledge the financial support from the Department of Science and Technology 
(DST), India, under the SERC project sanction grant No. SR/S2/HEP-29/2007. One of us (SU)
also acknowledge the financial support from CSIR, India, under SRF scheme.

\appendix

\section{  Mathematical details of Abelian 3-form gauge theory} 
\subsection{Extended BRST transformation of fields}
\begin{eqnarray}
&&s_b B_{\mu\nu\eta} = L_{\mu\nu\eta},\ \ s_b {\bar B}_{\mu\nu\eta}= L_{\mu\nu\eta}-
(\partial_\mu c_{\nu\eta} -\partial_\mu \bar c_{\nu\eta}+\partial_\nu c_{\eta\mu} -\partial_\nu \bar c_{\eta\mu} +
\partial_\eta c_{\mu\nu}-\partial_\eta \bar c_{\mu\nu}),\nonumber\\
&&s_b c_{\mu\nu} =  M_{\mu\nu},\ \ s_b\bar c_{\mu\nu} = M_{\mu\nu}-(\partial_\mu \beta_\nu -\partial_\mu 
\bar\beta_\nu - \partial_\nu \beta_\mu +\partial_\nu \bar\beta_\mu ),\
s_b \tilde c_{\mu\nu} = \tilde M_{\mu\nu},\nonumber\\
&& s_b\bar {\tilde c}_{\mu\nu}  =  \tilde M_{\mu\nu}-B_{\mu\nu} 
+\bar B_{\mu\nu},\ \
s_b B_{\mu\nu}  =  N_{\mu\nu},\ \ s_b\bar B_{\mu\nu} =N_{\mu\nu},\ \
s_b \tilde B_{\mu\nu} = \tilde N_{\mu\nu},\nonumber\\
&& s_b\bar {\tilde B}_{\mu\nu} =\tilde N_{\mu\nu} -(\partial_\mu f_\nu
-\partial_\mu\bar f_\nu -\partial_\nu f_\mu +\partial_\nu \bar f_\mu ),
\ s_b\beta_\mu  =  O_\mu,\   s_b\bar\beta_\mu = O_\mu -\partial_\mu c_2 +\partial_\mu \bar c_2, 
\nonumber\\
&&s_b\tilde \beta_\mu = \tilde O_\mu,\ \ s_b\bar{\tilde\beta}_\mu = \tilde O_\mu -\tilde F_\mu +\bar {\tilde F}_\mu, 
\ 
s_bF_\mu  =  P_\mu,\ \ s_b\bar F_\mu = P_\mu +\partial_\mu B -\partial_\mu \bar B, 
\nonumber\\
&&s_b\tilde F_\mu = \tilde P_\mu,\ \ s_b\bar{\tilde F}_\mu = \tilde P_\mu, \ \
s_bf_\mu = Q_\mu,\ \ s_b\bar f_\mu = Q_\mu, \ s_b\tilde f_\mu = \tilde Q_\mu,
\nonumber\\
 &&s_b\bar{\tilde f}_\mu = \tilde Q_\mu -\partial_\mu B_1 +\partial_\mu \bar B_1, 
\
s_b c_2  = R,\ \ s_b\bar c_2 =R,\
 s_b \tilde c_2 =\tilde R,\nonumber\\
&& s_b\bar {\tilde c}_2 =\tilde R -B_2 +\bar B_2, 
\
 s_b c_1 = S,\ \ s_b\bar c_1 =S +B -\bar B, 
\ \
 s_b \tilde c_1=\tilde S,\nonumber\\
 &&s_b\bar {\tilde c}_1 =\tilde S -B_1 +\bar B_1, 
\ \
s_b\phi_\mu  = T_\mu,\ \ s_b\bar{\phi}_\mu = T_\mu -f_\mu +\bar f_\mu,\
s_bB = U,\nonumber\\ 
&&s_b\bar B =U,\ 
 s_bB_1 =V,\ \ s_b\bar B_1 =V, \
s_bB_2  = W,\ \ s_b\bar B_2 =W,\nonumber\\
&&s_b \Omega =0, \label{bs1}
\end{eqnarray}
where $\Omega\equiv [L_{\mu\nu\eta}, M_{\mu\nu},  \tilde M_{\mu\nu}, N_{\mu\nu},  \tilde N_{\mu\nu}, O_\mu, \tilde 
O_\mu, P_\mu, \tilde P_\mu, Q_\mu, \tilde Q_\mu, R, \tilde R,  S, \tilde S, T_\mu, U, V, W]$.
\subsection{Extended BRST transformation of antifields}
\begin{eqnarray}
&&s_b B_{\mu\nu\eta}^\star =l_{\mu\nu\eta},\ \ 
s_b c_{\mu\nu}^\star = m_{\mu\nu},\ \
s_b \tilde{c}_{\mu\nu}^\star = \bar{m}_{\mu\nu},\ \
s_b B_{\mu\nu}^\star = n_{\mu\nu}, \nonumber\\
&&s_b \beta_\mu^\star = o_\mu, \ \
s_b \tilde {\beta}_\mu^\star = \bar o_\mu, \ \ 
s_b  {F_\mu}^\star = p_\mu, 
s_b \tilde{F_\mu}^\star = \bar p_\mu, \ \
s_b  {f_\mu}^\star = q_\mu,\nonumber\\
&&s_b \tilde{f_\mu}^\star = \bar q_\mu, \ \ 
s_b  {c_2}^\star = r,\ \
s_b \tilde{c_2}^\star = \bar r, \ \ 
s_b  {c_1}^\star = s,\ \
s_b \tilde{c_1}^\star = \bar s,\nonumber\\
&&s_b  {\phi_\mu}^\star = t_\mu,\  
s_b B^\star = u,\  
s_b  B_1^\star = v,\  
s_b B_2^\star  =  w,\ s_b \tilde{B}_{\mu\nu}^\star = \bar n_{\mu\nu},\ s_b \Lambda  =0,\nonumber\\
&& \Lambda \equiv l_{\mu\nu\eta}, m_{\mu\nu}, \bar{m}_{\mu\nu}, n_{\mu\nu},  \bar n_{\mu\nu},   o_
\mu, \bar o_\mu,
p_\mu,  \bar p_\mu, q_\mu, \bar q_\mu, r, \bar r, s, \bar s,  t_\mu, u, v, w.
\label{bs2}
\end{eqnarray}
\subsection{Superfields for the extended  BRST invariant theory} 
\begin{eqnarray} 
&&{\cal B}_{\mu\nu\eta}(x, \theta ) = B_{\mu\nu\eta} (x) +\theta L_{\mu\nu\eta},
 \ {\cal C}_{\mu\nu}(x, \theta )  = c_{\mu\nu} (x) +\theta M_{\mu\nu},\nonumber\\
&&\bar{{\cal B}}_{\mu\nu\eta}(x, \theta ) = \bar{B}_{\mu\nu\eta} (x) +\theta (L_{\mu\nu\eta}-
(\partial_\mu c_{\nu\eta} -\partial_\mu \bar c_{\nu\eta}+\partial_\nu c_{\eta\mu} -\partial_\nu \bar c_{\eta\mu} +
\partial_\eta c_{\mu\nu}-\partial_\eta \bar c_{\mu\nu}) ),\nonumber\\
&&\bar{\cal C}_{\mu\nu}(x, \theta ) = \bar c_{\mu\nu} (x) +\theta ( M_{\mu\nu}-(\partial_\mu \beta_\nu -\partial_\mu 
\bar\beta_\nu - \partial_\nu \beta_\mu +\partial_\nu \bar\beta_\mu )),\nonumber\\
&&\tilde{\cal C}_{\mu\nu}(x, \theta ) =\tilde c_{\mu\nu} (x) +\theta \tilde M_{\mu\nu}, \
\bar{\tilde{\cal C}}_{\mu\nu}(x, \theta )  =  \bar {\tilde c}_{\mu\nu} (x) +\theta ( \tilde M_{\mu\nu}-B_{\mu\nu} 
+\bar B_{\mu\nu}),\nonumber\\ 
&&{\cal B}_{\mu\nu}(x, \theta ) = B_{\mu\nu} (x) +\theta N_{\mu\nu}, \ \
\bar{\cal B}_{\mu\nu}(x, \theta )  =  \bar B_{\mu\nu} (x) +\theta N_{\mu\nu},\nonumber\\
&&\tilde{\cal B}_{\mu\nu}(x, \theta ) = \tilde B_{\mu\nu} (x) +\theta \tilde N_{\mu\nu},\ {\cal B}_{\mu}(x, \theta ) 
=  \beta_{\mu} (x) +\theta O_{\mu},\nonumber\\
&&\bar{\tilde{\cal B}}_{\mu\nu}(x, \theta ) = \bar {\tilde B}_{\mu\nu} (x) +\theta (\tilde N_{\mu\nu} 
-(\partial_\mu f_\nu -\partial_\mu\bar f_\nu -\partial_\nu f_\mu +\partial_\nu \bar f_\mu )),\nonumber\\
&&\bar{\cal B}_{\mu}(x, \theta ) = \bar \beta_{\mu } (x) +\theta (O_\mu -\partial_\mu c_2 +\partial_\mu \bar c_2),
\nonumber\\
&&\tilde{\cal B}_{\mu}(x, \theta ) =\tilde \beta_{\mu} (x) +\theta \tilde O_{\mu}, \
\bar{\tilde{\cal B}}_{\mu}(x, \theta ) = \bar {\tilde\beta}_{\mu } (x) +\theta (\tilde O_\mu -\tilde F_\mu +\bar 
{\tilde F}_\mu),\nonumber\\
&&{\cal F}_{\mu}(x, \theta ) = F_{\mu} (x) +\theta P_{\mu},\ 
\bar{\cal F}_{\mu}(x, \theta )  =  \bar F_{\mu } (x) +\theta (P_\mu +\partial_\mu B -\partial_\mu \bar B),
\nonumber\\
&&\tilde{\cal F}_{\mu}(x, \theta ) =\tilde F_{\mu} (x) +\theta \tilde P_{\mu},\ 
\bar{\tilde{\cal F}}_{\mu}(x, \theta )  =  \bar {\tilde F}_{\mu } (x) +\theta \tilde P_{\mu},\nonumber\\
&&{\bf f}_{\mu}(x, \theta ) = f_{\mu} (x) +\theta Q_{\mu},\ \
\bar{\bf f}_{\mu}(x, \theta )  =  \bar f_{\mu} (x) +\theta  Q_{\mu},
\nonumber\\
&&\tilde{\bf f}_{\mu}(x, \theta ) = \tilde f_{\mu} (x) +\theta \tilde Q_{\mu},\  
\bar{\tilde{\bf f}}_{\mu}(x, \theta )  =  \bar {\tilde f}_{\mu} (x) +\theta (\tilde Q_\mu -\partial_\mu B_1 
+\partial_\mu \bar B_1),\nonumber\\
&&{\cal C}_{2}(x, \theta ) = c_{2} (x) +\theta R,\ \
\bar{\cal C}_{2}(x, \theta )  =  \bar c_{2} (x) +\theta R,
\nonumber\\
&&\tilde{\cal C}_{2}(x, \theta ) = \tilde c_{2} (x) +\theta \tilde R,\ 
\bar{\tilde{\cal C}}_{2}(x, \theta )  =  \bar {\tilde c}_{2} (x) +\theta (\tilde R -B_2 +\bar B_2),\nonumber\\
&&{\cal C}_{1}(x, \theta ) = c_{1} (x) +\theta S,\ 
\bar{\cal C}_{1}(x, \theta )  =  \bar c_{1} (x) +\theta (S +B -\bar B),
\nonumber\\
&&\tilde{\cal C}_{1}(x, \theta ) = \tilde c_{1} (x) +\theta \tilde S,\ 
\bar{\tilde{\cal C}}_{1}(x, \theta )  =  \bar {\tilde c}_{1} (x) +\theta (\tilde S -B_1 +\bar B_1),\nonumber\\
&&{ \Phi}_{\mu}(x, \theta )= \phi_\mu (x) +\theta T_\mu,\ 
\bar{  \Phi}_{\mu}(x, \theta )  =  \bar \phi_\mu (x) +\theta ( T_\mu -f_\mu +\bar f_\mu),
\nonumber\\
&&{ \cal B}(x, \theta ) =B(x) +\theta U,\ \
\bar{  \cal B}(x, \theta )  =  \bar B (x) +\theta U,
\nonumber\\
&&{ \cal B}_1(x, \theta ) =B_1(x) +\theta V,\ \
\bar{  \cal B}_1(x, \theta )  =  \bar B_1 (x) +\theta V,
\nonumber\\
&&{ \cal B}_2(x, \theta )=B_2(x) +\theta W,\ \
\bar{  \cal B}_2(x, \theta )  =  \bar B_2 (x) +\theta W,\nonumber\\
&&\bar{\cal B}_{\mu\nu\eta}^\star = B_{\mu\nu\eta}^\star  +\theta L_{\mu\nu\eta}, \ \
\bar{{\cal C}}_{\mu\nu}^\star = c_{\mu\nu}^\star  +\theta M_{\mu\nu}, \ \
\bar{\tilde{\cal C}}_{\mu\nu}^\star  =  \tilde c_{\mu\nu}^\star  +\theta \tilde M_{\mu\nu},\nonumber\\
&&\bar{{\cal B}}_{\mu\nu}^\star = B_{\mu\nu}^\star  +\theta N_{\mu\nu},\ \
\bar{\tilde{\cal  B}}_{\mu\nu}^\star  =  \tilde B_{\mu\nu}^\star  +\theta \tilde N_{\mu\nu},\ \
\bar{{\cal B}}_{\mu}^\star = \beta_{\mu}^\star  +\theta O_{\mu},\nonumber\\
&&\bar{\tilde{\cal B}}_{\mu}^\star =\tilde \beta_{\mu}^\star  +\theta\tilde O_{\mu},\ \ 
\bar{{\cal F}}_{\mu}^\star = F_{\mu}^\star  +\theta P_{\mu},\ \
\bar{\tilde{\cal F}}_{\mu}^\star  =  \tilde F_{\mu}^\star  +\theta \tilde P_{\mu},\nonumber \\
&&\bar{{\bf f}}_{\mu}^\star=f_{\mu}^\star  +\theta Q_{\mu},\ \
\bar{\tilde{\bf f}}_{\mu}^\star  = \tilde f_{\mu}^\star  +\theta \tilde Q_{\mu},\ \
\bar{{\cal C}}_{2}^\star = c_{2}^\star  +\theta R,\nonumber\\ 
&&\bar{\tilde{\cal C}}_{2}^\star =\tilde c_{2}^\star  +\theta \tilde R,\ \
\bar{{\cal C}}_{1}^\star = c_{1}^\star  +\theta S,\ \
\bar{\tilde{\cal C}}_{1}^\star  = \tilde c_{1}^\star  +\theta \tilde S,\nonumber\\
&&\bar{  \Phi}_{\mu}^\star =\phi_\mu^\star  +\theta T_\mu,\ \
\bar{ {\cal B}}^\star  = B^\star  +\theta U,\ \
\bar{{\cal B}}_{1}^\star =B_{1}^\star  +\theta V,\ \
\bar{{\cal B}}_{2}^\star =B_{2}^\star  +\theta W. \label{supb}
\end{eqnarray}
\subsection{Superfields for both extended BRST and anti-BRST invariant theory}
\begin{eqnarray}  
{\cal B}_{\mu\nu\eta}(x, \theta, \bar\theta ) &=& B_{\mu\nu\eta} (x) +\theta L_{\mu\nu\eta}
+\bar \theta(B_{\mu\nu\eta}^\star + 
\partial_\mu \tilde c_{\nu\eta}-\partial_\mu \bar{\tilde c}_{\nu\eta}+\partial_\nu \tilde c_{\eta\mu}
-\partial_\nu \bar{\tilde c}_{\eta\mu}  +\partial_\eta 
\tilde c_{\mu\nu}\nonumber\\
&-&\partial_\eta \bar{\tilde c}_{\mu\nu}) 
+\theta\bar\theta (l_{\mu\nu\eta} +\partial_\mu B_{\nu\eta} -\partial_\mu \bar B_{\nu\eta}
+\partial_\nu B_{\eta\mu} -\partial_\nu \bar B_{\eta\mu} + \partial_\eta B_{\mu\nu} -\partial_\eta \bar B_{\mu\nu}),
\nonumber\\
  {\cal C}_{\mu\nu}(x, \theta, \bar\theta ) & =& c_{\mu\nu} (x) +\theta M_{\mu\nu} +\bar \theta (c_{\mu\nu}^\star +
\tilde B_{
\mu\nu} -\bar{\tilde B}_{\mu\nu} )+\theta\bar\theta (m_{\mu\nu} +\partial_\mu f_\nu -\partial_\mu \bar f_\nu -
\partial_\nu f_\mu \nonumber\\
&+&\partial_\nu \bar f_\mu ),\nonumber\\
\bar{{\cal B}}_{\mu\nu\eta}(x, \theta, \bar\theta ) &=& \bar{B}_{\mu\nu\eta} (x) +\theta (L_{\mu\nu\eta}-
(\partial_\mu c_{\nu\eta} -\partial_\mu \bar c_{\nu\eta}+\partial_\nu c_{\eta\mu} -\partial_\nu \bar c_{\eta\mu} +
\partial_\eta c_{\mu\nu}-\partial_\eta \bar c_{\mu\nu}) )\nonumber\\
&+&\bar\theta B_{\mu\nu\eta}^\star +\theta\bar\theta l_{\mu\nu\eta},\nonumber\\
\bar{\cal C}_{\mu\nu}(x, \theta, \bar\theta ) &=& \bar c_{\mu\nu} (x) +\theta ( M_{\mu\nu}-(\partial_\mu \beta_\nu -
\partial_\mu 
\bar\beta_\nu - \partial_\nu \beta_\mu +\partial_\nu \bar\beta_\mu ))+\bar\theta c_{\mu\nu}^\star
 + \theta\bar\theta   m_{\mu\nu},\nonumber\\
\tilde{\cal C}_{\mu\nu}(x, \theta, \bar\theta ) &=&\tilde c_{\mu\nu} (x) +\theta \tilde M_{\mu\nu}+\bar\theta(\tilde 
c_{\mu\nu}^
\star +\partial_\mu \tilde\beta_\nu -\partial_\mu 
\bar{\tilde\beta}_\nu -\partial_\nu \tilde\beta_\mu +\partial_\nu \bar{\tilde\beta}_\mu ) 
\nonumber\\
&+&\theta\bar\theta ( \bar m_{\mu\nu}+\partial_\mu \tilde F_\nu -\partial_\mu\bar{\tilde F}_\nu -\partial_\nu \tilde 
F_\mu +\partial_\nu\bar{\tilde F}_\mu), 
\nonumber\\
\bar{\tilde{\cal C}}_{\mu\nu}(x, \theta, \bar\theta ) & =&  \bar {\tilde c}_{\mu\nu} (x) +\theta ( \tilde M_{\mu\nu
}-B_{\mu\nu} 
+\bar B_{\mu\nu}) +\bar\theta  {\tilde c}_{\mu\nu}^\star +\theta\bar\theta \bar m_{\mu\nu},\nonumber\\ 
{\cal B}_{\mu\nu}(x, \theta, \bar\theta ) &=& B_{\mu\nu} (x) +\theta N_{\mu\nu}+\bar\theta (B_{\mu\nu}^\star + 
\partial_\mu\tilde f_\nu -\partial_\mu\bar{\tilde f}_\nu
-\partial_\nu\tilde f_\mu +\partial_\nu\bar{\tilde f}_\mu )
+\theta\bar\theta n_{\mu\nu},\nonumber\\
\bar{\cal B}_{\mu\nu}(x, \theta, \bar\theta ) & =&  \bar B_{\mu\nu} (x) +\theta N_{\mu\nu}+\bar\theta B_{\mu\nu}^\star
+\theta\bar\theta n_{\mu\nu},\nonumber\\
\tilde{\cal B}_{\mu\nu}(x, \theta, \bar\theta ) &=& \tilde B_{\mu\nu} (x) +\theta \tilde N_{\mu\nu}+\bar\theta \tilde
 B_{\mu\nu}^\star
+\theta\bar\theta \bar n_{\mu\nu},
\nonumber\\
 {\cal B}_{\mu}(x, \theta, \bar\theta ) 
&=&  \beta_{\mu} (x) +\theta O_{\mu}+\bar\theta ( \beta_{\mu}^\star +F_\mu -\bar F_\mu )+\theta\bar\theta (
o_\mu -\partial_\mu B_2 +\partial_\mu \bar B_2),
\nonumber\\
\bar{\tilde{\cal B}}_{\mu\nu}(x, \theta, \bar\theta ) &=& \bar {\tilde B}_{\mu\nu} (x) +\theta (\tilde N_{\mu\nu} 
-(\partial_\mu f_\nu -\partial_\mu\bar f_\nu -\partial_\nu f_\mu +\partial_\nu \bar f_\mu ))+\bar\theta {\tilde B}_{
\mu\nu}^\star
+\theta\bar\theta \bar n_{\mu\nu},\nonumber\\
\bar{\cal B}_{\mu}(x, \theta, \bar\theta ) &=& \bar \beta_{\mu } (x) +\theta (O_\mu -\partial_\mu c_2 +\partial_\mu 
\bar c_2)+\bar\theta \beta_{\mu }^\star +\theta\bar\theta o_\mu,
\nonumber\\
\tilde{\cal B}_{\mu}(x, \theta, \bar\theta ) &=& \tilde \beta_{\mu} (x) +\theta \tilde O_{\mu}+\bar\theta (
{\tilde\beta}_{\mu}^\star +\partial_\mu \tilde c_2 -\partial_\mu \bar{\tilde c}_2)
+\theta\bar\theta ( \bar o_\mu +\partial_\mu B_2 -\partial_\mu \bar B_2),\nonumber\\
\bar{\tilde{\cal B}}_{\mu}(x, \theta, \bar\theta )& =& \bar {\tilde\beta}_{\mu} (x) +\theta (\tilde O_\mu -\tilde 
F_\mu +\bar {\tilde F}_\mu)+\bar\theta {\tilde\beta}_{\mu}^\star +\theta\bar\theta \bar o_\mu,\nonumber\\
{\cal F}_{\mu}(x, \theta, \bar\theta ) &=& F_{\mu} (x) +\theta P_{\mu}+\bar\theta F_{\mu}^\star 
+\theta\bar\theta p_\mu,\nonumber\\ 
\bar{\cal F}_{\mu}(x, \theta, \bar\theta ) & =&  \bar F_{\mu} (x) +\theta (P_\mu +\partial_\mu B -\partial_\mu 
\bar B)+\bar\theta F_{\mu}^\star +\theta\bar\theta p_\mu,\nonumber\\
\tilde{\cal F}_{\mu}(x, \theta, \bar\theta ) &=& \tilde F_{\mu} (x) +\theta \tilde P_{\mu}+\bar\theta (\tilde 
F_{\mu}^\star -\partial_\mu B_2 +
\partial_\mu\bar B_2)+\theta\bar\theta \bar p_\mu,\nonumber\\ 
\bar{\tilde{\cal F}}_{\mu}(x, \theta, \bar\theta ) &= &\bar {\tilde F}_{\mu} (x) +\theta \tilde P_{\mu}
+\bar\theta {\tilde F}_{\mu}^\star +\theta\bar\theta \bar p_\mu,\nonumber\\
{\bf f}_{\mu}(x, \theta, \bar\theta ) &=& f_{\mu} (x) +\theta Q_{\mu}+\bar\theta (f_{\mu}^\star -\partial_\mu B_1 +
\partial_\mu\bar B_1)+\theta\bar\theta  q_\mu,\nonumber\\
\bar{\bf f}_{\mu}(x, \theta, \bar\theta )  &=&  \bar f_{\mu} (x) +\theta  Q_{\mu}+\bar\theta f_{\mu}^\star
+\theta\bar\theta q_\mu,\nonumber\\
\tilde{\bf f}_{\mu}(x, \theta, \bar\theta ) &=& \tilde f_{\mu} (x) +\theta \tilde Q_{\mu}+\bar\theta
{\tilde f}_{\mu}^\star +
\theta\bar\theta \bar q_\mu,\nonumber\\  
\bar{\tilde{\bf f}}_{\mu}(x, \theta, \bar\theta )  &=&  \bar {\tilde f}_{\mu} (x) +\theta (\tilde Q_\mu -
\partial_\mu B_1 +\partial_\mu \bar B_1)+\bar\theta {\tilde f}_{\mu}^\star +\theta\bar\theta \bar q_\mu,\nonumber\\
{\cal C}_{2}(x, \theta, \bar\theta ) &=& c_{2} (x) +\theta R+\bar\theta (c_{2}^\star +B-\bar B)+\theta\bar\theta r,
\nonumber\\
\bar{\cal C}_{2}(x, \theta, \bar\theta ) & = & \bar c_{2} (x) +\theta R+\bar\theta c_{2}^\star +\theta\bar\theta r,
\nonumber\\
\tilde{\cal C}_{2}(x, \theta, \bar\theta ) &=& \tilde c_{2} (x) +\theta \tilde R+\bar\theta \tilde c_{2}^\star
+\theta\bar\theta \bar r,\nonumber\\ 
\bar{\tilde{\cal C}}_{2}(x, \theta, \bar\theta )  &= & \bar {\tilde c}_{2} (x) +\theta (\tilde R -B_2 +\bar B_2)+
\bar\theta {\tilde c}_{2}^\star +\theta\bar\theta \bar r,\nonumber\\
{\cal C}_{1}(x, \theta, \bar\theta ) &=& c_{1} (x) +\theta S+\bar\theta (c_{1}^\star -B_1 +\bar B_1)+\theta\bar\theta
 s,\nonumber\\ 
\bar{\cal C}_{1}(x, \theta, \bar\theta )  &= & \bar c_{1} (x) +\theta (S +B -\bar B)+\bar\theta c_{1}^
\star +\theta\bar\theta s,\nonumber\\
\tilde{\cal C}_{1}(x, \theta, \bar\theta ) &=& \tilde c_{1} (x) +\theta \tilde S+\bar\theta (\tilde c_{1}^\star -B_2+
\bar B_2)+\theta\bar\theta \bar s,
\nonumber\\ 
\bar{\tilde{\cal C}}_{1}(x, \theta, \bar\theta ) &= & \bar {\tilde c}_{1} (x) +\theta (\tilde S -B_1 +\bar B_1)+
\bar\theta {\tilde c}_{1}^\star +\theta\bar\theta \bar s,\nonumber\\
{ \Phi}_{\mu}(x, \theta, \bar\theta ) &=& \phi_\mu (x)  +\theta T_\mu +\bar\theta (\phi_\mu^\star +\tilde f_\mu -
\bar{\tilde f}_\mu )+\theta\bar\theta (t_\mu +\partial_\mu B_1 -\partial_\mu \bar B_1),\nonumber\\
\bar{  \Phi}_{\mu}(x, \theta, \bar\theta )  &= & \bar \phi_\mu (x) +\theta ( T_\mu -f_\mu +\bar f_\mu)+\bar\theta 
 \phi_\mu^\star
+\theta\bar\theta t_\mu,\nonumber\\
{ \cal B}(x, \theta, \bar\theta ) &=&B(x) +\theta U+\bar\theta B^\star+\theta\bar\theta u,\nonumber\\
\bar{  \cal B}(x, \theta, \bar\theta ) & = & \bar B (x) +\theta U+\bar\theta B^\star+\theta\bar\theta u,
\nonumber\\
{ \cal B}_1(x, \theta, \bar\theta ) &=&B_1(x) +\theta V+\bar\theta B_1^\star+\theta\bar\theta v,\nonumber\\
\bar{  \cal B}_1(x, \theta, \bar\theta )  &=&  \bar B_1 (x) +\theta V+\bar\theta B_1^\star+\theta\bar\theta v,
\nonumber\\
{ \cal B}_2(x, \theta, \bar\theta ) &=&B_2(x) +\theta W+\bar\theta B_2^\star+\theta\bar\theta w,\nonumber\\
\bar{  \cal B}_2(x, \theta, \bar\theta ) & =&  \bar B_2 (x) +\theta W +\bar\theta B_2^\star +\theta\bar\theta w. 
\label{superf}
\end{eqnarray}
Form the above relations, we calculate   
\begin{eqnarray}
 \frac{1}{2}\frac{\delta}{\delta\bar\theta}\frac{\delta}{\delta\theta}\bar {\cal B}_{\mu\nu\eta} \bar {\cal B}^{
\mu\nu\eta} 
&=& l_{\mu\nu\eta}\bar B^{\mu\nu\eta} -B_{\mu\nu\eta}^\star (L^{\mu\nu\eta} -\partial^\mu c^{\nu\eta} +\partial^\mu 
\bar c^{\nu\eta} -\partial^\nu c^{\eta\mu}+\partial^\nu \bar c^{\eta\mu}\nonumber\\
&-&\partial^\eta c^{ \mu\nu}+\partial^\eta \bar 
c^{ \mu\nu}),\nonumber\\
  \frac{\delta}{\delta\bar\theta}\frac{\delta}{\delta\theta}\bar{\tilde {\cal C}}_{\mu\nu}\bar {\cal C}^{
\mu\nu}
&=& \bar m_{\mu\nu}\bar c^{\mu\nu} +m_{\mu\nu}\bar {\tilde c}^{\mu\nu} +\tilde c_{\mu\nu}^\star (
M^{\mu\nu} -\partial^\mu \beta^\nu +\partial^\mu \bar \beta^\nu +\partial^\nu \beta^\mu -\partial^\nu \bar \beta^\mu)
\nonumber\\
&+& c_{\mu\nu}^\star (\tilde M^{\mu\nu} -B^{\mu\nu} +\bar B^{\mu\nu}),
\nonumber\\
 \frac{1}{2}\frac{\delta}{\delta\bar\theta}\frac{\delta}{\delta\theta} \bar{\tilde {\cal B}}_{\mu\nu} \bar{\tilde {
\cal B}}^{\mu\nu}
&=&  n_{\mu\nu}\bar B^{\mu\nu}-B_{\mu\nu}^\star N^{\mu\nu}, 
\nonumber\\
 \frac{1}{2}\frac{\delta}{\delta\bar\theta}\frac{\delta}{\delta\theta}\bar{\cal B}_{\mu\nu} \bar{\cal B}^{\mu\nu}
&=& \bar n_{\mu\nu}\bar {\tilde B}^{\mu\nu} -\tilde B_{\mu\nu}^\star (\tilde N^{\mu\nu} -\partial^\mu f^\nu +
\partial^\mu\bar f^\nu +\partial^\nu f^\mu -\partial^\nu \bar f^\mu ),
\nonumber\\
 \frac{1}{2}\frac{\delta}{\delta\bar\theta}\frac{\delta}{\delta\theta}\bar{\cal B}_{\mu } \bar{\cal B}^{\mu}
&=& o_\mu\bar\beta^\mu -\beta_\mu^\star (O^\mu -\partial^\mu c_2 +\partial^\mu \bar c_2),
\nonumber\\
 \frac{1}{2}\frac{\delta}{\delta\bar\theta}\frac{\delta}{\delta\theta}\bar{\tilde {\cal B}}_{\mu } 
\bar{\tilde{\cal B}}^{\mu}
&=& \bar o_\mu\bar{\tilde \beta}^\mu -\tilde\beta_\mu^\star (\tilde O^\mu -\tilde F^\mu +\bar{\tilde F}^\mu),
\nonumber\\
  \frac{\delta}{\delta\bar\theta}\frac{\delta}{\delta\theta}\bar{\tilde {\cal F}}_{\mu } \bar{\cal F}^{\mu} 
&=& \bar p_\mu\bar F^\mu +p_\mu \bar {\tilde F}^\mu +\tilde F_\mu^\star (P^\mu +\partial^\mu B -\partial ^\mu \bar 
B) +F_\mu^\star \tilde P^\mu,\nonumber\\
  \frac{\delta}{\delta\bar\theta}\frac{\delta}{\delta\theta}\bar{\tilde {\bf f}}_{\mu } \bar{\bf f}^{\mu} 
&=& \bar q_\mu\bar f^\mu +q_\mu\bar{\tilde f}^\mu +\tilde f_\mu^\star Q^\mu +f_\mu^\star (
\tilde Q^\mu -\partial^\mu B_1+\partial^\mu\bar B_1),\nonumber\\
 \frac{\delta}{\delta\bar\theta}\frac{\delta}{\delta\theta} \bar{\tilde {\cal C}}_{2 } \bar
{\cal C}_2
&=&  \bar r\bar c_2 +r\bar {\tilde c}_2 +\tilde c_2^\star R +c_2^\star (\tilde R -B_2 +\bar B_2),\nonumber\\
  \frac{\delta}{\delta\bar\theta}\frac{\delta}{\delta\theta}\bar{\tilde {\cal C}}_{1} \bar{{\cal C}}_1
&=& \bar s \bar{ c}_1 +\tilde c_1^\star ( S +B -\bar B ) +s\bar{\tilde c}_1 +c_1^\star
(\tilde S-B_1 +\bar B_1),\nonumber\\
 \frac{1}{2}\frac{\delta}{\delta\bar\theta}\frac{\delta}{\delta\theta}\bar{\Phi}_{\mu } \bar{ \Phi}^{\mu}
&=& t_\mu\bar \phi^\mu -\phi_\mu^\star (T^\mu -f^\mu +\bar f^\mu ),\nonumber\\
 \frac{1}{2}\frac{\delta}{\delta\bar\theta}\frac{\delta}{\delta\theta}\bar {\cal B}\bar {\cal B}
&=&u\bar B -B^\star U,\nonumber\\
 \frac{1}{2}\frac{\delta}{\delta\bar\theta}\frac{\delta}{\delta\theta}\bar {\cal B}_1\bar {\cal B}_1 
&=&  v\bar B_1 -B_1^\star V,\nonumber\\
 \frac{1}{2}\frac{\delta}{\delta\bar\theta}\frac{\delta}{\delta\theta}\bar {\cal B}_2\bar {\cal B}_2
&=&w\bar B_2 -B_2^\star W. 
\end{eqnarray}


\begin{thebibliography}{0}
\bibitem{green} M. Green, J. Schwarz and E. Witten, {\it{Superstring theory}},  (Cambridge 
Univ. Press, 1987). 
\bibitem{pol} J. Polchinski, {\it{ String theory}},  (Cambridge 
Univ. Press, 1998).
\bibitem{h} M. B. Green, J. H. Schwarz and E. Witten, {\it{Superstring Theory}}, (Cambridge 
Univ. Press, 1987).
\bibitem{i} J. Polchinski, {\it{String Theory}} (Cambridge Univ. Press, 1998). 
\bibitem{a} M. Kalb and P. Ramond, {\it{Phys. Rev.}} {\bf{D 9}}, 2273 (1974).
\bibitem{b} F. Lund and T. Regge, {\it{Phys. Rev.}} {\bf{D 14}}, 1524 (1976).
\bibitem{c} M. Sato and S. Yahikozawa, {\it{Nucl. Phys.}} {\bf{B 436}}, 100 (1995).
\bibitem{d} A. Sugamoto, {\it{Phys. Rev.}} {\bf{D 19}}, 1820 (1979). 
\bibitem{e} R. L. Davis and E. P. S. Shellard, {\it{Phys. Lett.}}, {\bf{B 214}}, 219 (1988).
\bibitem{g} A. Salam and E. Sezgin, {\it{Supergravities in Diverse Dimensions}} 
(North-Hplland and World Scientific, 1989).
\bibitem{degu} Shinichi Deguchi, Tomoaki Mukai and Tadahito Nakajima, {\it Phys. Rev.} {\bf D 59}, 065003 (1999).
\bibitem{brst} C. Becchi, A. Rouet and R. Stora, {\it Annals Phys.} {\bf{98}} 287 (1974). 
\bibitem {tyu}I. V. Tyutin, LEBEDEV-{\bf 75-39} (1975).
\bibitem{ht} M. Henneaux and C. Teitelboim, {\it{ Quantization of gauge
systems}} (Princeton, USA: Univ. Press, 1992).
\bibitem{wei} S. Weinberg, {\it{ The quantum theory of fields, Vol-II: Modern
applications}} (Cambridge, UK Univ. Press, 1996).
\bibitem{bv} I. A. Batalin and G. A. Vilkovisky, {\it{ Phys. Lett.}}{\bf{ B 102}}, 27 (1981).
\bibitem{bv1} I. A. Batalin and G. A. Vilkovisky, {\it{Phys. Rev.}} {\bf{D 28}},2567 (1983); 
Erratum ibid {\bf{D 30}}, 508 (1984).
\bibitem{bv2} I. A. Batalin and G. A. Vilkovisky, {\it{Phys. Lett.}} {\bf{ B 120}} 166 (1983).
\bibitem{subm}  S. Upadhyay and B. P. Mandal, arXiv:1112.0422[hep-th].
\bibitem{ad} J. Alfaro and P. H. Damgaard, {\it Phys. Lett.} {\bf B 222}, 425 (1989); J. 
Alfaro, P. H.
Damgaard , J. I. Latorre and D. Montano, {\it Phys. Lett.} {\bf B 233}, 153 (1989); J. Alfaro
and P. H. Damgaard, {\it Nucl. Phys.} {\bf B 404}, 751 (1993).
\bibitem{ba} N. R.F. Braga and A. Das, {\it Nucl. Phys.} {\bf B 442}, 655 (1995).
\bibitem{fk} M. Faizal, M. Khan, {\it Eur. Phys. J.} {\bf C 71}, 1603 (2011).
\bibitem{sm} S. Upadhyay and B. P. Mandal, {\it Mod. Phys. Lett.} {\bf A 25 }, 3347 (2010).
\bibitem{al} J. Alfaro and P.H. Damgaard,  {\it Nucl. Phys.} {\bf B 404}, 751  (1993).
\bibitem{bt} L. Bonora and M. Tonin, {\it Phys. Lett.} {\bf B 98}, 48 (1981).
\bibitem{cf} G. Curci and R. Ferrari, {\it{ Phys. Lett.}}{\bf{ B 63}}, 91 (1976).
\bibitem{lm} L. Bonora and R. P. Malik, {\it J. Phys. A : Math. Theor. } {\bf 43}, 375403 (2010).
\end{thebibliography}
\end{document}